\documentclass[12pt]{article}
\pdfoutput=1
\usepackage[margin=3cm]{geometry}
\usepackage[utf8]{inputenc}
\usepackage{amsmath}
\usepackage{amssymb}
\usepackage{authblk} % for authors
\usepackage{xcolor}
\usepackage{physics}
\usepackage{graphicx} % for graphics in pdf_tex figures
\usepackage{float}
\usepackage{tikz}
\usetikzlibrary{calc}
\usepackage{comment}

\usepackage{acronym}
\newacro{EFT}[EFT]{effective field theory}
\newacro{AdS}[AdS]{anti-de Sitter}
\newacro{dS}[dS]{de Sitter}
\acrodefplural{EFT}{effective field theories}
\newacro{SUSY}[SUSY]{supersymmetry}
\newacro{KK}[KK]{Kaluza-Klein}

\usepackage{hyperref}
\usepackage{varioref}       
\usepackage{cleveref}     

\crefname{table}{table}{tables}
\Crefname{table}{Table}{Tables}
\crefname{figure}{figure}{figures}
\Crefname{figure}{Figure}{Figures}

\definecolor{tealblue}{rgb}{0.21, 0.56, 0.63}
\hypersetup{colorlinks=true,allcolors = tealblue,linktocpage=true}

\usepackage{pgf} % for pgf pics, might have to remove on publication if pgfs have to be prerendered

\usepackage[
        backend=bibtex,
        sorting=none,
        style=phys,
        eprint=true,
        doi=false,
        biblabel=brackets
        ]{biblatex}

\bibliography{bibliograph}

\newcommand{\ads}{\text{AdS}}

\usepackage[backgroundcolor=white!95!red, linecolor=cyan, bordercolor=cyan, textsize=footnotesize]{todonotes}

\newcommand{\commie}[1]{}

\numberwithin{equation}{section}
\numberwithin{table}{section}

\newenvironment{eqaed}
    {\begin{equation}
    \begin{aligned}
    }
    { 
    \end{aligned}
    \end{equation}
    \ignorespacesafterend
    }

\begin{document}

\title{Dynamical dark energy in 0'B braneworlds}
\date{}

\author{Ivano Basile$^a$\thanks{ibasile@mpp.mpg.de}}
\author{Alessandro Borys$^{a,b}$\thanks{alessandro.borys@dfa.unict.it}}
\author{Joaquin Masias$^a$\thanks{jmasias@mpp.mpg.de}}

\affil{${}^a$\emph{Max-Planck-Institut f\"ur Physik (Werner-Heisenberg-Institut)}\\ \emph{Boltzmannstraße 8, 85748 Garching, Germany}}

\affil{${}^b$\emph{Department of Physics, University of Catania}\\ \emph{Via Santa Sofia 64, I-95123 Catania, Italy}}

\maketitle

\begin{abstract}
    
We build a novel realization of dark bubble cosmology in non-supersymmetric string theory. Among the simplest models in ten dimensions, the type 0'B orientifold is the unique option which yields a scale-separated construction. The resulting setting produces a logarithmically varying dynamical dark energy, reflecting its holographic counterpart in terms of running gauge couplings. We analyze in detail the phenomenological consequences of the model for particle physics, inflation and late-time cosmology. We find that, although particle physics may be consistently realized, neither early-time nor late-time are observationally viable.
    
\end{abstract}

\thispagestyle{empty}

\newpage

\tableofcontents

\thispagestyle{empty}

\newpage

\pagenumbering{arabic}

\section{Introduction} \label{sec:introduction}

One of the main goals of string phenomenology is to build a realistic model of our universe from string theory. 
The standard approach is to begin with the simplest sectors of the theory, where extra dimensions exist, and study time-independent\footnote{Time-dependent compactifications may also be an option, although they complicate matters significantly \cite{Brahma:2020tak, Bernardo:2021rul, Chakravarty:2024pec}.} compactifications to extract low-energy four-dimensional \acp{EFT}. While the string landscape is vast and contains non-geometric and non-supersymmetric phases, research has largely focused on geometric settings in which \ac{SUSY} is broken within the \ac{EFT}. This approach then faces the challenge of stabilizing moduli, controlling \ac{SUSY}-breaking effects and produce a realistic description of particle physics. See \cite{VanRiet:2023pnx, Cicoli:2023opf} for recent reviews of this overarching paradigm.

Furthermore, our universe is undergoing a phase of accelerated expansion, which appears well-described by a (quasi-)\ac{dS} geometry. However, persistent difficulties in constructing a (meta)stable \ac{dS} vacuum in string theory \cite{Bena:2023sks} have led to conjecture that no such vacuum can exist, at least toward the boundary of moduli space \cite{obied2018sitterspaceswampland}. On the other hand, \ac{AdS} vacua are common in string theory, but there are reasons to believe that, unless they are supersymmetric, they are unstable and must decay either perturbatively or non-perturbatively \cite{Ooguri:2016pdq}. All these difficulties notwithstanding, several efforts to construct \ac{dS} vacua in explicit string constructions are underway \cite{Kachru:2003aw, Demirtas:2021nlu, Demirtas:2021ote, McAllister:2024lnt}. Since compactifications lead to \emph{bona fide} \acp{EFT} of gravity, it is natural to apply swampland \cite{Palti:2019pca, van_Beest_2022, agmon2023lecturesstringlandscapeswampland} reasonings to them. Combining these with observational data recently led to the dark dimension proposal \cite{Montero:2022prj}, whose realizations \cite{Blumenhagen:2022zzw, Basile:2024lcz}, phenomenological necessities \cite{Anchordoqui:2025nmb} and consequences \cite{Gonzalo:2022jac, Anchordoqui:2022txe, Anchordoqui:2022tgp, Anchordoqui:2023oqm, Obied:2023clp, Vafa:2024fpx, Antoniadis:2024sfa, Anchordoqui:2024xvl, Vafa:2025nst} are being actively explored.

An alternative approach to describe our universe without dealing with the compactification of these extra dimensions (or non-geometric sectors) and a soft breaking of \ac{SUSY} are the so-called braneworld scenarios, in which our universe is embedded in a higher-dimensional spacetime (the ``bulk'') and the Standard Model lives on a brane. Settings of this type have been proposed for other reasons as well \cite{Randall_1999, Randall:1999ee, Giddings:2000mu, Karch:2000ct, Dvali:2000hr}, and they can be combined with compactification into a powerful toolbox for particle physics \cite{Aldazabal:2000cn, Lust:2004ks} potentially together with cosmology \cite{Antoniadis:2019rkh}.
 
In this respect, the recently developed ``dark bubble'' scenario \cite{Banerjee:2018qey, Banerjee_2019, Banerjee:2020wix, Banerjee:2021yrb, Danielsson:2021tyb, Banerjee:2022ree, Danielsson:2022lsl, Danielsson:2022fhd, Basile:2023tvh, Banerjee:2023uto, danielsson2024chargednariaiblackholes} combines together string theory and cosmology in a beautiful and novel interplay. In this scenario, our universe rides a four-dimensional bubble in a higher-dimensional spacetime, which crucially must contain an \ac{AdS} factor. The latter is unstable and decays non-perturbatively, nucleating bubbles of lower energy. The dark bubble proposal therefore involves inducing an effective four-dimensional gravity on a suitable brane which mediates the decay from a non-supersymmetric false vacuum to a lower energy configuration. The latter may itself be unstable, or it may be supersymmetric \cite{Banerjee:2018qey}. Due to the relative abundance of \ac{AdS} constructions in string theory, some explicit realizations of the dark bubble scenario have been proposed \cite{Basile:2020mpt, Danielsson:2022lsl}. Compared to the more popular appraoch based on compactifications, the dark bubble scenario has several perks, but also aspects which could be considered flaws.

The main advantages of the dark bubble are:

\begin{itemize}
    \item Breaking \ac{SUSY} is inherently necessary, and is the very first requirement;
    \item A positive (effective) dark energy on the braneworld is automatically produced, and follows from the decay process in the bulk;
    \item A separation of scales \cite{Coudarchet:2023mfs} between the brane and the bulk can be achieved rather easily, by concocting brane configurations which are extremal to leading order;
    \item The difficulties entailed by moduli stabilization are avoided, since the ambient spacetime is qualitatively similar to the familiar $\ads_5 \times S^5$ supported by fluxes. Moreover, the propagation of long-range forces in the bulk does not translate straightforwardly to braneworld physics;
    \item Producing the required gauge groups and chiral fermions without superpartners is feasible.
\end{itemize}

While the above list can surely appeal to the interested phenomenologist, one must bear in mind issues of similar importance:

\begin{itemize}
    \item The dark bubble is not described by a \emph{bona fide} four-dimensional quantum \ac{EFT} coupled to gravity. Instead, at least classically, the four-dimensional Einstein equations at low energies arise from non-localized degrees of freedom. This may lead to some phenomenological inconsistencies given the robustness of the \ac{EFT} paradigm, although on the flip side it would provide characteristic predictions outside of it;
    \item Relatedly, the couplings of matter and gauge fields to the effective induced gravity are not \emph{a priori} guaranteed to be those predicted by standard \ac{EFT}. As we shall see, there may be substantial differences in some cases;
    \item The absence of a direct \ac{EFT} description means that swampland constraints are at best indirectly related to the construction, and at worst wholly inapplicable. If the dark bubble were to provide (infinitely?) many realistic models of particle physics and cosmology, the deep lessons of finiteness \cite{Hamada:2021yxy, Kim:2024hxe, Delgado:2024skw} and constraining power \cite{Montero:2020icj, Bedroya:2021fbu} provided by the swampland program would be wasted when applied to phenomenology.
\end{itemize}

In this paper we put forth a novel realization of the dark bubble in string theory where \ac{SUSY} is absent from the outset. Our construction is based on the unique non-tachyonic ten-dimensional orientifold of the type 0B string, dubbed ``type 0'B'' or Sagnotti model \cite{Sagnotti:1995ga, Sagnotti:1996qj}. This setting is particularly simple from the stringy viewpoint: \ac{SUSY} is entirely absent and there is no need to introduce extra ingredients to break it. Moreover, the ambient \ac{AdS} spacetime can be realized by a simple configuration of parallel D3-branes. This leads to a four-dimensional cosmology which automatically incorporates unitary gauge groups and chiral fermions. In fact, among the simplest realizations of this type, which include the ones of \cite{Basile:2020mpt}, this is the unique option which yields a parametric scale separation between brane and bulk physics. We examine this model in detail, obtaining a parameter space where realistic particle physics may be viable. However, when turning to cosmological aspects, it turns out that it is observationally excluded. In contrast with the construction of \cite{Danielsson:2022lsl}, the type 0'B dark bubble has a characteristic feature: a logarithmically varying dynamical dark energy, which stems from the quantum properties of the worldvolume gauge theory on D3-branes which allows for scale separation in the first place. It turns out that this particular type of dynamical dark energy produces a blue-shifted power spectrum, which excludes an inflationary realization of this model. Similarly, the predictions for late-time cosmology are in conflict with Standard Model couplings. All in all, the model appears not to be phenomenologically viable.

The paper is structured as follows. In \cref{sec:non-susy_strings} we provide a brief overview of non-supersymmetric string theory in ten dimensions, focusing on the aspects we need. In \cref{sec:review_dark bubblec} we introduce the dark bubble scenario from a bottom-up perspective, mentioning some salient features of the top-down construction of \cite{Danielsson:2022fhd}. In this context, in \cref{EM} we extend the analysis of braneworld electromagnetic fields in \cite{Basile:2023tvh} to the case relevant for the type 0'B model. In \cref{sec:0'B_dS} we construct the dark bubble scenario in type 0'B string theory using D3-branes, obtaining a logarithmically varying dynamical dark energy as a general prediction of models of this type. We prove a no-go result regarding its effective description as four-dimensional scalar fields, showing that only phantom scalars can recover this behavior. In \cref{sec:pheno} we delve into the overlap of this model with experimental bounds, starting from particle physics in \cref{sec:scales_bounds} and then proceeding to early- and late-time cosmology in \cref{sec:cosmo}. We collect our findings and provide some concluding remarks in \cref{sec:conclusions}.

\section{A brief review of non-supersymmetric strings} \label{sec:non-susy_strings}

Spacetime \ac{SUSY} plays an important role in string theory and related subjects, but whether it is ultimately instrumental for its consistency is not understood. Moreover, the lack of positive empirical evidence for \ac{SUSY} at the level of particle physics ($\approx 13$  TeV) indicates that fundamental \ac{SUSY}, if at all present, is broken at some unknown scale. A deeper understanding of the subtle issues related to \ac{SUSY} breaking in string theory is the first step to achieve a more complete picture of its underlying foundational principles as well as more realistic phenomenological models. In the context of this paper, the dark bubble scenario inherently requires some form of \ac{SUSY} breaking. In particular, we shall focus on realizations within string models where \ac{SUSY} is either broken at the string scale or absent altogether, at least at the perturbative level. In this section, we will briefly discuss these non-supersymmetric constructions. These settings arise in Ramond-Neveu-Schwarz (RNS) superstring theory in ten spacetime dimensions, and they comprise the only models which do not contain tachyons in their spectra. The main focus of this paper is one of these models, dubbed type 0'B or Sagnotti model.

\subsection{Tachyon-free vacuum amplitudes}

Let us begin by constructing the relevant ten-dimensional string models starting from the one-loop vacuum amplitudes. Differently from (perturbative) field theory, the consistency of GSO projections and the uniqueness of interactions entail that one-loop vacuum amplitudes contain a great deal of information about string vacua and their excitations. Although one-loop vacuum amplitudes \emph{per se} are numerical-valued loop integrals of single-particle partition functions (over the fundamental domain $\mathcal{F}$ of complex structures $\tau$ of the worldsheet torus), the latter can be written in a way which manifests the physical features of the spectrum\footnote{This can be seen more explicitly using character-valued \cite{Schellekens:1986xh} or even representation-valued \cite{Hanany:2010da} partition functions.}. This can be achieved expressing one-loop integrands via the characters $(O_{2n},V_{2n},S_{2n},C_{2n})$ of the level-1 affine $\mathrm{so}(2n)$ algebra, associated to its four conjugacy classes. This algebra appears as the transverse isometry group to the string worldsheet in flat spacetime. In the heterotic case, it also appears in the internal sector and encodes the gauge structure of the theory. The four characters are \cite{Angelantonj:2002ct}
\begin{align}
    O_{2n}&\equiv\frac{\vartheta^n\begin{bmatrix}
0\\
0
\end{bmatrix}(0|\tau)+\vartheta^n\begin{bmatrix}
0\\
1/2
\end{bmatrix}(0|\tau)}{2\eta^n(\tau)}, \nonumber\\ \nonumber
     V_{2n}&\equiv\frac{\vartheta^n\begin{bmatrix}
0\\
0
\end{bmatrix}(0|\tau)-\vartheta^n\begin{bmatrix}
0\\
1/2
\end{bmatrix}(0|\tau)}{2\eta^n(\tau)}, \\ \nonumber
      S_{2n}&\equiv\frac{\vartheta^n\begin{bmatrix}
1/2\\
0
\end{bmatrix}(0|\tau)+i^{-n}\vartheta^n\begin{bmatrix}
1/2\\
1/2
\end{bmatrix}(0|\tau)}{2\eta^n(\tau)}, \\ 
       C_{2n}&\equiv\frac{\vartheta^n\begin{bmatrix}
1/2\\
0
\end{bmatrix}(0|\tau)-i^{-n}\vartheta^n\begin{bmatrix}
1/2\\
1/2
\end{bmatrix}(0|\tau)}{2\eta^n(\tau)},
\end{align}
where $\eta(\tau)\equiv q^{\frac{1}{24}}\Pi_{n=1}^\infty(1-q^n)$ is the Dedekind $\eta$ function and the Jacobi $\vartheta$ functions are defined as
\begin{equation}
    \vartheta\begin{bmatrix}
\alpha\\
\beta
\end{bmatrix}(z|\tau)\equiv\sum_{n\in\mathrm{Z}}q^{\frac{1}{2}(n+\alpha)^2}\mathrm{e}^{2\pi i(n+\alpha)(z-\beta)}.
\end{equation}
These characters encode degeneracies of states in various spacetime (and/or gauge) representations, and can be used to express worldsheet partition functions. Taking into account open descendants, in ten-dimensional Minkowski spacetime there are three non-tachyonic models without (unbroken) spacetime \ac{SUSY}. They can all be built from projections of closed-string models.

\begin{itemize}
\item \textbf{Sugimoto model}: this model is an orientifold projection of the type IIB string. The (halved) torus amplitude
\begin{equation}
     \mathcal{T}=\frac{1}{2}\int_{\mathcal{F}}\frac{d^2\tau}{\tau^6_2}\frac{(V_8-S_8)\overline{(V_8-S_8)}}{|\eta(\tau)|^{16}}
    \end{equation}
is to be supplemented with the contributions pertaining to the open-string and unoriented sectors, which are associated to the Klein bottle, the annulus and the M\"{o}bius strip. These involve a Chan-Paton degeneracy $N$ and a reflection coefficient $\epsilon$, and take the form \cite{Angelantonj:2002ct}
\begin{align}
     \mathcal{K}&=\frac{1}{2}\int_{0}^\infty\frac{d\tau_2}{\tau^6_2}\frac{(V_8-S_8)(2i\tau_2)}{\eta^8(2i\tau_2)}, \\ 
     \mathcal{A}&=\frac{N^2}{2}\int_{0}^\infty\frac{d\tau_2}{\tau^6_2}\frac{(V_8-S_8)(i\frac{\tau_2}{2})}{\eta^8(\frac{i\tau_2}{2})}, \\ 
     \mathcal{M}&=\frac{\varepsilon N}{2}\int_{0}^\infty\frac{d\tau_2}{\tau^6_2}\frac{(\hat{V_8}+\hat{S_8})(\frac{i\tau_2}{2}+\frac{1}{2})}{\hat{\eta}^8(\frac{i\tau_2}{2}+\frac{1}{2})}.
\end{align}
The divergences in these amplitudes are interpreted in the loop channel of open/closed duality as tadpoles the NS-NS and R-R sectors, whose cancellation is essential for consistency and requires $N=32$, $\varepsilon=1$. 
This leads to a $USp(32)$ gauge group\footnote{See \cite{Larotonda:2024thv} for a recent analysis of its global structure.}. However, because the resulting O9-plane has positive tension and charge, the latter is cancelled by $\overline{D9}$-branes \cite{Sugimoto:1999tx}, while the dynamical NS-NS tadpole remains. As a result, \ac{SUSY} is broken at the string scale\footnote{More precisely, \ac{SUSY} is preserved in the closed-string sector, but it is non-linearly realized in the open-sector via "Brane Supersymmetry Breaking" \cite{Antoniadis:1999xk, Angelantonj:1999jh, Aldazabal:1999jr, Angelantonj:1999ms}, as reflected in the presence of a Goldstone fermion in the spectrum. The realization of the super-Higgs mechanism in this context remains elusive \cite{Dudas:2000nv, Dudas:2000ff, Dudas:2001wd}.}. This dynamical tadpole is sourced by the residual combined tension $T=64T_\text{D9}$ of the D-branes and O-plane, which is encoded in the Einstein-frame runaway exponential potential
\begin{equation}\label{eq:sugimoto_pot}
    T\int d^{10}x\sqrt{-g}e^{\gamma\phi}, \hspace{20 pt} \gamma=\frac{3}{2} \, .
\end{equation}

\item \textbf{Sagnotti model}: starting from the type 0B string, whose torus amplitude reads
\begin{equation}
\mathcal{T}_{0'B}=\int_{\mathcal{F}}\frac{d^2\tau}{\tau^6_2}\frac{O_8\overline{O_8}+V_8\overline{V_8}+S_8\overline{S_8}+C_8\overline{C_8}}{|\eta(\tau)|^{16}} \, , \label{eq:torus}
\end{equation}
there are several orientifold projections available \cite{Sagnotti:1995ga, Sagnotti:1996qj}, one of which is tachyon-free. This choice yields the $U(32)$ type 0'B model, and involves adding to (half of) \cref{eq:torus} the open-sector amplitudes
\begin{align}
 \mathcal{K}_\text{0'B}&=\frac{1}{2}\int_{0}^\infty\frac{d\tau_2}{\tau^6_2}\left(-O_8+V_8+S_8-C_8\right) , \\ 
     \mathcal{A}_\text{0'B} &=\int_{0}^\infty\frac{d\tau_2}{\tau^6_2} \left(n\overline{n}V_8-\frac{n^2+\overline{n}^2}{2}C_8\right) , \\ 
     \mathcal{M}_\text{0'B} &=\int_{0}^\infty\frac{d\tau_2}{\tau^6_2}\frac{n+\overline{n}}{2}\hat{C_8} \, . 
\end{align}
Tadpole cancellation fixes $n=\overline{n}=32$ and a $U(32)$ gauge group, whose overall $U(1)$ is pseudo-anomalous and undergoes a Stuckelberg mechanism together with the R-R axion. The corresponding O9-plane has vanishing tension, and therefore the resulting Einstein-frame exponential potential
\begin{equation}
    T\int d^{10}x\sqrt{-g}e^{\gamma\phi}, \hspace{20 pt} \gamma=\frac{3}{2}
\end{equation}
if half of that of \cref{eq:sugimoto_pot}.

\item \textbf{SO(16)$\times$SO(16) model}: starting from the torus amplitude of the $E_8\times E_8$ superstring,
\begin{equation}
\mathcal{T}_{HE}=\int_{\mathcal{F}}\frac{d^2\tau}{\tau^6_2}\frac{(V_8-S_8)\overline{(O_{16}+S_{16})}^2}{|\eta(\tau)|^{16}} \, ,
\end{equation}
the projection onto the states with even total fermion number produces the unique non-supersymmetric heterotic string without tachyons in ten dimensions\footnote{Recently, the classification of chiral conformal field theories of central charge 16 confirmed this result \cite{BoyleSmith:2023xkd}.}, whose gauge algebra is that of $SO(16) \times SO(16)$\footnote{Once more, in this paper we are not careful about the global structure of gauge groups. In this case, this aspect was discussed in \cite{McInnes:1999va} and is relevant for global anomalies \cite{Basile:2023knk}, gravitational solitons \cite{Kneissl:2024zox} and CHL-like constructions \cite{Nakajima:2023zsh, Angelantonj:2024jtu}.} Due to modular invariance, the projected torus amplitude is to be supplemented with its images under $S$ and $T$ modular transformations. The final amplitude reads
\begin{align}
\mathcal{T}_{SO(16)\times SO(16)}=\int_{\mathcal{F}}\frac{d^2\tau}{\tau^6_2}\frac{1}{{|\eta(\tau)|^{16}}}&\bigg[O_8\overline{(V_{16}C_{16}+C_{16}V_{16})}\\ \nonumber
&+V_8\overline{(O_{16}O_{16}+S_{16}S_{16})} \\ \nonumber
&-S_8\overline{(O_{16}S_{16}+S_{16}O_{16})} \\ \nonumber
&-C_8\overline{(V_{16}V_{16}+C_{16}C_{16})}\bigg] \, . 
\end{align}
Although there is no tree-level tadpole, the one-loop vacuum energy does not vanish \cite{Dixon:1986iz, AlvarezGaume:1986jb}. In the string-frame low-energy effective action it appears as a cosmological constant, and thus in the Einstein frame it corresponds to a runaway exponential potential
\begin{equation}
    T\int d^{10}x\sqrt{-g}e^{\gamma\phi} \, , \hspace{20 pt} \gamma=\frac{5}{2} \, .
\end{equation}
\end{itemize}

All in all, the low-energy manifestation of dynamical tadpoles, at least for our purposes, can be encompassed by the same type of exponential potential for the dilaton in all three cases.

\subsection{Brane content and low-energy effective description}

The D-brane content of the orientifold models we reviewed was studied in \cite{Sugimoto:1999tx, Dudas:2001wd}. Focusing on branes carrying integral Ramond-Ramond charge, the Sugimoto model contains charged D1-branes and D5-branes, as in the type I string, while the Sagnotti model contains charged D$p$-branes with $p$ odd, as in the type IIB string.
At leading order, the D$p$-D$p$ interaction between charged branes vanishes, but the presence of the background $\overline{\text{D}9}$-branes and $\text{O}9$-plane brings along additional contributions \cite{Antonelli:2019nar, Basile_2021}. 

With the construction of a dark bubble scenario in mind, we now briefly review the low-energy effective description of the three non-supersymmetric models introduced in the preceding section. The notation is taken from \cite{Mourad:2016xbk, Antonelli:2019nar, Basile_2021}. Insofar as weakly curved and weakly coupled backgrounds and processes can reliably account for tadpoles, the low-energy physics is described the effective actions of the form
\begin{equation}
    S=\frac{1}{2\kappa_D^2}\int d^Dx\sqrt{-g}\left(R-\frac{4}{D-2}(\partial \phi)^2-V(\phi)-\frac{f(\phi)}{2(p+2)!}H_{p+2}^2\right), \label{eq:lowenergyS}
\end{equation}
where here and in the following $\kappa^2_D \equiv 8\pi G_D$ is Newton's constant in $D$-dimensions. The bosonic fields include a dilaton $\phi$ and a $(p+2)$-form field-strength $H_{p+2}=B_{p+1}$. Yang-Mills fields are also present, but we shall not need them in the following. In the relevant string models $D=10$, and
\begin{equation}
    V(\phi)=Te^{\gamma\phi}, \hspace{20 pt} f(\phi)=e^{\alpha\phi} \, .
\end{equation}
The low-energy dynamics of the orientifold models with D1 and/or D5 fluxes is described by the Einstein-frame parameters $p=1$, $\gamma= \frac{3}{2}$, $\alpha=1$, while for the heterotic model $p=1$, $\gamma=\frac{5}{2}$, $\alpha=-1$. Finally, and most importantly for this work, the Sagnotti model features a self-dual five-form field strength, as in type IIB, for which $f(\phi)=1$.

\begin{figure}[ht!]
    \centering \includegraphics[width=0.8\textwidth]{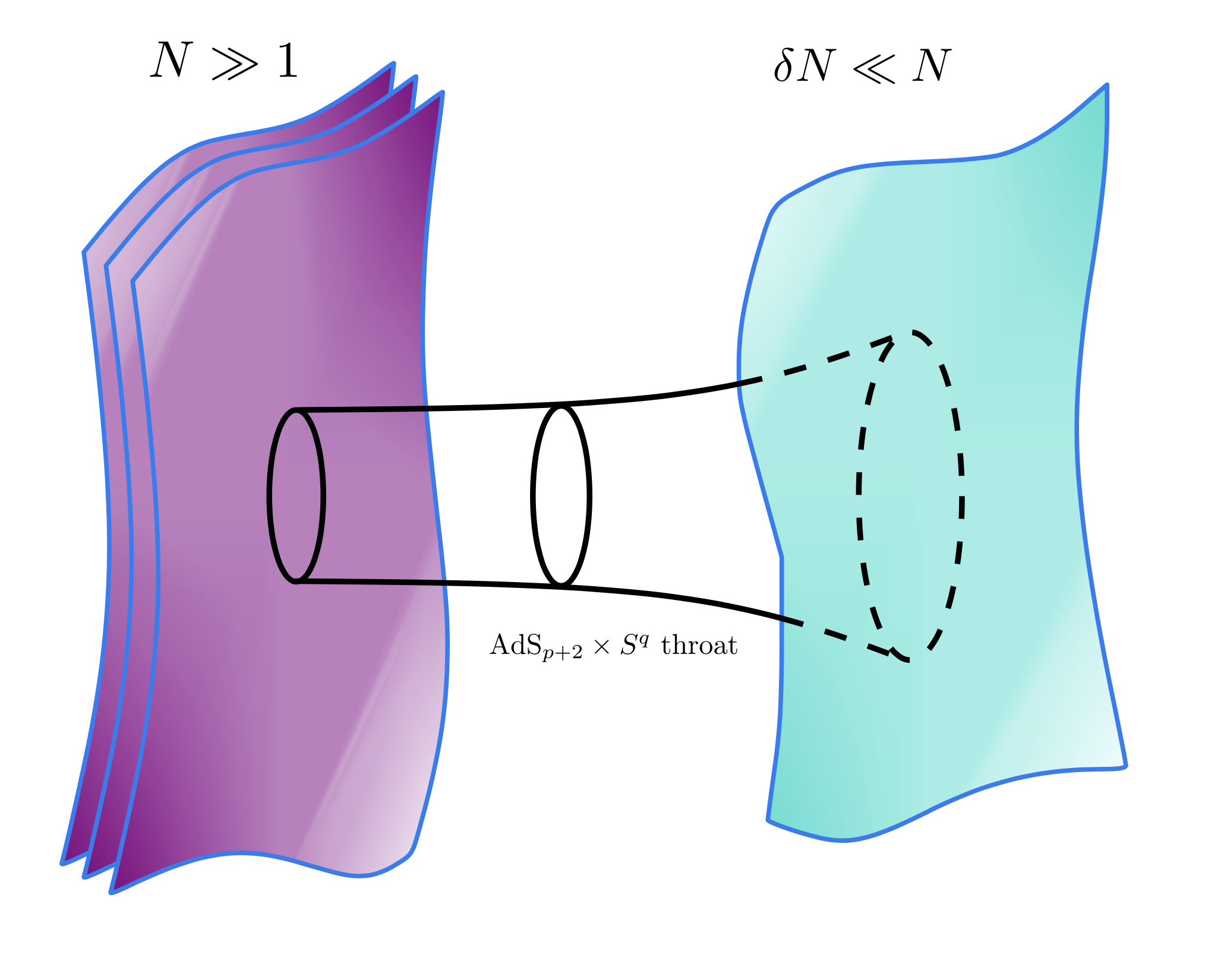}
    \caption{Schematic representation of the interaction between a heavy stack of $N \gg 1$ branes and $\delta N \ll N$ probe branes. The heavy stack sources the $AdS\times S$ throat probed by the light stack. Taken from \cite{Basile_2021}.}
    \label{fig:el}
\end{figure}

The field equations derived from \cref{eq:lowenergyS} read 
\begin{align}
    R_{MN}&=\tilde{T}_{MN} \, , \nonumber \\
    \Box\phi-V'(\phi)-\frac{f'(\phi)}{2(p+2)!}H_{p+2}^2&=0 \, , \\ \nonumber
    d*(f(\phi)H_{p+2})&=0 \, ,
\end{align}
where the trace-reversed stress-energy tensor is
\begin{equation}
   \tilde{T}_{MN}=T_{MN}-\frac{1}{D-2}T^A_Ag_{MN} \, , \hspace{20 pt} T_{MN}\equiv\frac{\delta S_{matter}}{\delta g^{MN}} \, .
\end{equation}
For the action in \cref{eq:lowenergyS}, this translates into
\begin{align}
    \tilde{T}_{MN}&=\frac{4}{D-2}\partial_M\partial_N\phi+\frac{f(\phi)}{2(p+1)!}(H_{p+2}^2)_{MN} \\
    &+\frac{g_{MN}}{D-2}\left(V-\frac{p+1}{2(p+2)!}f(\phi)H_{p+2}^2\right).
\end{align}
The interactions between branes can be studied in various complementary regimes \cite{Basile_2021}. For our purposes it is sufficient to employ the effective description above, whereby probe branes interact with the background sourced by heavy branes. The upshot of this analysis \cite{Antonelli:2019nar} (see also \cite{Mourad:2021roa, Mourad:2024dur, Mourad:2024mpg}), as depicted in \cref{fig:el}, is that near-horizon throats of extremal branes are \ac{AdS} (or quasi-\ac{AdS} \cite{Dudas:2000sn, Angelantonj:1999qg, Angelantonj:2000kh}) geometries, akin to the familiar supersymmetric $\ads_5 \times S^5$ of type IIB. Extremal probe branes of equal dimension are strictly repelled by the throat, realizing the weak gravity conjecture \cite{ArkaniHamed:2006dz} due to the \ac{SUSY}-breaking ingredients. Similarly, the NS-NS interactions in the presence of at least one uncharged stack are repulsive or attractive depending on their dimensions, matching the behavior expected from a worldsheet analysis \cite{Basile_2021}.

\subsection{Probe potentials and Weak Gravity}\label{sec:probe}

As we have remarked, the weak-gravity repulsion between charged branes is instrumental in building a dark bubble scenario in the settings at stake. However, as we shall now review, for all the available branes except D3-branes in the Sagnotti model the resulting potentials are not phenomenologically viable. In order to see this, the starting point is a general string-frame worldvolume action of the form
\begin{equation}
    S_p=-T_p\int d^{p+1}\zeta\sqrt{-j^*g_S} \, e^{-\sigma\phi}+\mu_p\int B_{p+1} \, ,
\end{equation}
where $j$ is the embedding of the worldvolume coordinates $\zeta$ in space-time. Its Einstein-frame counterpart thus reads
\begin{equation}
    S_p=-T_p\int d^{p+1}\zeta\sqrt{-j^*g_E}e^{\left(\frac{2(p+1)}{D-2}-\sigma\right)\phi}+\mu_p\int B_{p+1} \, , \label{eq:Eframe}
\end{equation}
where $\sigma=1,2$ for D-branes and NS5-branes respectively.

In order to compute the probe potential in the $\ads_{p+2} \times S^q$ throats obtained in \cite{Mourad:2016xbk, Antonelli:2019nar}, we work in Poincar\'{e} coordinate, where the Einstein-frame metric of the $AdS\times S$ throat takes the form 
\begin{equation}
    ds^2=\frac{L^2}{z^2}(dz^2+dx^2_{1,p})+R^2d\Omega_q^2 \, .
\end{equation}
We place the probe stack parallel to the throat, according to the worldvolume embedding $j: x^\mu=\zeta^\mu, z=Z(\zeta), \theta^i=\theta^i_0,$ with $\theta_0^i$ fixed coordinates on $S^q$. As a result, the potential felt by rigid probes simplifies to
\begin{align}
    V_\text{probe}(Z)&=\tau_p\left(\frac{L}{Z}\right)^{p+1}\left[1-\frac{cLg_s^{\frac{\alpha}{2}}}{p+1}\frac{\mu_p}{T_p}\right] \\ \nonumber
    &\equiv\tau_p\left(\frac{L}{Z}\right)^{p+1}\left[1-v_0\,\frac{\mu_p}{T_p}\right] ,
\end{align}
with the dressed tension $\tau_p\equiv T_p g_s^{-\frac{\alpha}{2}}$. In the above expression the extremality factor $v_0 > 1$ turns out to be a fixed constant \cite{Antonelli:2019nar}, showing repulsion of like-charge branes. Although realizing the weak gravity conjecture in this setting is a non-trivial consistency check for non-supersymmetric strings, an order-one extremality factor translates into dark bubble models \emph{without scale separation} \cite{Basile:2020mpt}. In other words, a low-energy observer on the bubble would be able to detect ten-dimensional bulk physics. In order to circumvent this problem, the branes must be mutually extremal to leading order in this analysis, so that the leading $v_0 = 0$ is corrected by loop effects to $v_0 \ll 1$. This will generate the required scale separation between the bulk and the braneworld. In order to see how this mechanism takes place, in the following section we shall discuss the main features and implications of the dark bubble scenario from a bottom-up perspective. We will then move on to the stringy realization in the Sagnotti type 0'B model, whose D3-branes provide the only option to realize a scale-separate dark bubble scenario among the ones we discussed in the \cref{sec:non-susy_strings}. This construction also results in a four-dimensional quasi-\ac{dS} cosmology, whose phenomenological viability we shall discuss in detail in \cref{sec:pheno}.

\section{The dark bubble scenario} \label{sec:review_dark bubblec}

The dark bubble model \cite{Banerjee:2018qey, Banerjee_2019, Banerjee:2020wix, Banerjee:2021yrb, Danielsson:2021tyb, Banerjee:2022ree, Danielsson:2022lsl, Danielsson:2022fhd, Basile:2023tvh, Banerjee:2023uto, danielsson2024chargednariaiblackholes} is an inside-outside construction where two $\ads_5$ vacua with different cosmological constants are separated by a brane\footnote{This model markedly differs from the Randall-Sundrum  (RS) construction \cite{Randall_1999}, in which two insides of the bubble are glued together.}. This construction modifies the way in which a form of induced gravity is realized on the brane, since there is no compactification. Accordingly, matter is also delocalized as endpoints of strings stretching along the radial dimension. 
    
Lower-dimensional observers are confined on a (3+1)-dimensional brane and perceive Einstein gravity as an effective 4d theory. Indeed, the 4d Einstein equations follow from the junction condition across the brane via the Gauss-Codazzi relation. The brane geometry is sourced by the energy-momentum tensor of the brane itself (which for empty branes acts as a cosmological constant) and by contributions from the higher-dimensional bulk geometry. Upon adding matter to the brane, the full back-reacted solution needs to be considered. This leads to a net \textit{positive} energy density, taking into account the extrinsic curvature through the junction condition. Stretched strings pulling on the brane look like particles in 4d, whose mass is related to the string tension by the bulk geometry in such a way that only \ac{AdS} produces a constant mass \cite{Basile:2020mpt}. The bending of the brane due to the energy density of a string cloud (corresponding to dust in 4d) reproduces standard FRLW cosmology.

As a final introductory remark, we notice that the dynamics of the dark bubble is perfectly consistent with Vilenkin's tunneling proposal in quantum cosmology, with the bubble nucleation in 5d AdS spacetime identified with a 4d Big Bang event. The amplitude of bubble nucleation in 5d matches Vilenkin's tunneling amplitude in the 4d description \cite{Danielsson:2021tyb}.

\begin{figure}[H]
\centering
\includegraphics[width=0.7\textwidth]{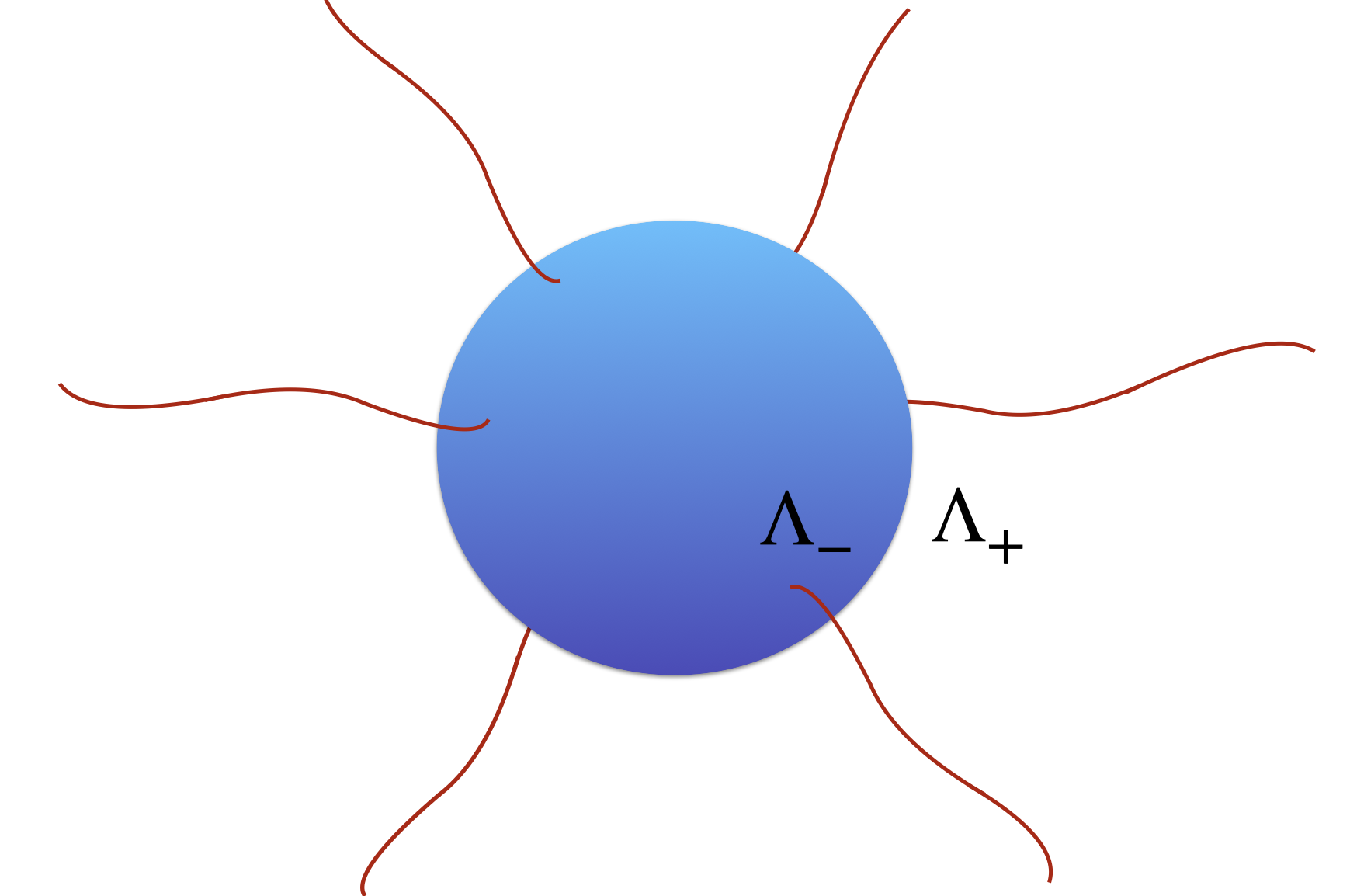}
      \caption{\textit{Pictorial representation of the dark bubble, with an inside/outside construction such that $\Lambda_-<\Lambda_+$. Strings extending all the way to infinity correspond to massive particles on the braneworld.}}
      \label{fig:bubbles}
\end{figure}
    
\subsection{Gravity on the braneworld}

As outlined above, the stress-energy tensor on the brane sources a jump in the extrinsic curvature. The Israel junction conditions \cite{Israel:1966rt} imply that the tension $\sigma$ of the brane be given by
\begin{equation}
        \sigma=\frac{3}{8\pi G_5}\left(\sqrt{k_-^2+\frac{1+\dot a^2}{a^2}}-\sqrt{k_+^2+\frac{1+\dot a^2}{a^2}}\right),
    \end{equation}
where the cosmological constants inside and outside the shell are given by $\Lambda_\pm=-6k_\pm^2=-6/L_\pm^2$ and satisfy $\Lambda_-<\Lambda_+<0$ (or $k_->k_+$). The vacuum with higher energy can then decay through the nucleation of a spherical bubble, which depending on the specific construction may be a Coleman-de Luccia or Brown-Teitelboim instanton.

In terms of the proper time $\tau$ on the brane located at $r=a(\tau)$ in global \ac{AdS} coordinates, the induced metric on the brane has the FLRW form
\begin{equation}
    ds^2_\text{brane}=-d\tau^2+a(\tau)^2 d\Omega_3^2 \, .
\end{equation}
When $k_- - k_+ \ll k_\pm$ the tension is slightly subcritical, and the scale factor satisfies
\begin{equation}
    \frac{\dot a^2}{a^2}\sim -\frac{1}{a^2}+\frac{8\pi G_4}{3}\Lambda_4 \, ,
\end{equation}
where 
\begin{equation}
        \Lambda_4=\frac{3(k_--k_+)}{8\pi G_5}-\sigma \, , \qquad G_4 = \frac{2k_-k_+}{k_--k_+} \, G_5 
\end{equation}
play the role of (positive) cosmological constant and 4d Newton constant. 

$\sigma_\text{cr} \equiv \frac{3}{8\pi G_5}(k_--k_+)$ is the critical tension. Since the 4d cosmological constant is proportional to $\sigma_\text{cr} - \sigma$, we see how a slightly subcritical brane is essential to realize scale separation of the braneworld physics with respect to the bulk.

As discussed in \cite{Banerjee:2020wix}, matter and radiation in the 4d effective cosmology arise from string clouds and bulk black holes respectively. Indeed, a warp factor
\begin{equation}
   f(r)_\pm=1+k_\pm^2r^2-\frac{\kappa_5M_\pm}{3\pi^2 r^2}-\frac{\kappa_5 \alpha_\pm}{4\pi r},
\end{equation}
in global \ac{AdS} coordinates, with a discontinuity in the density of strings and black-hole mass, induces a scale factor driven by
\begin{equation}
        H^2=\frac{\dot a^2}{a^2}\sim -\frac{1}{a^2}+\frac{8\pi G_4}{3}\left[\Lambda_4+\frac{1}{2\pi^2a^4}\left(\frac{M_+}{k_+}-\frac{M_-}{k_-}\right)+ \frac{3}{8\pi a^3}\frac{\tau}{k_+}\right].
\end{equation}
All in all, we see that the mechanism which induces a 4d cosmology on the brane does not pertain to standard 4d \ac{EFT}. In other words, the physics is inherently higher-dimensional, and the separation of scales follows from the interplay of various higher-dimensional ingredients, rather than how it occurs in a standard compactification.

\subsection{Stringy constructions of dark bubble cosmology}

Shortly after its inception, top-down string constructions of the dark bubble scenario were sought. In the non-supersymmetric string models outlined in \cref{sec:non-susy_strings}, Freund-Rubin flux compactification leads to \ac{AdS} vacua \cite{Mourad:2016xbk} whose perturbative instabilities were studied in \cite{Basile:2018irz}. These settings, while promising in light of the probe-brane analysis of \cref{sec:probe}, ultimately turned out to produce braneworlds without scale separation \cite{Basile:2020mpt}. As outlined in \cref{sec:probe}, it became apparent that scale separation requires the brane to be extremal at the order at which the \ac{AdS} geometry appears, and receive subleading superextremal corrections.

Eventually, a type IIB embedding of the dark bubble model in string theory was concocted \cite{Danielsson:2022lsl}. The crucial ingredient in this construction is rotating branes in the internal dimension, which breaks \ac{SUSY} and generates a non-zero but parametrically small extremality parameter via subleading corrections to the worldvolume effective action. The probe limit is the large-$N$ limit of the background $\ads_5 \times S^5$ solution, corresponding to a large number $N$ of source branes.

The resulting physical scales have peculiar scalings with $N$, which we shall compare to our type 0'B construction in \cref{sec:0'B_dS}. The Hubble radius $R_H$, the \ac{AdS} curvature radius $L$, and the ten- and five-dimensional Planck lengths $l_{10}, l_5$ are expressed in terms of the four-dimensional Planck length $l_4$ according to
\begin{equation}\label{eq:iib_scalings}
    R_H\sim Nl_4, \hspace{10 pt} L\sim N^{\frac{1}{2}}l_4, \hspace{10 pt} l_{10}\sim N^{\frac{1}{4}}l_4, \hspace{10 pt} l_5\sim N^{-\frac{1}{6}}l_4 \, .
\end{equation}
Finally, the string length scale is $l_s\sim l_{10}/g_s^{\frac{1}{4}}$. Comparison with observation fixes $N \approx 10^{60}$ and $g_s \approx 10^{-2}$ \cite{Danielsson:2023alz}. As a result $l_s \approx 10$ TeV, an exciting prospect for accelerator experiments. Moreover $L \approx 10 \mu \text{m}$ is compatible with the predictions of the dark dimension scenario \cite{Montero:2022prj}, albeit with a completely different origin.

\subsection{Embedding electromagnetism in the bubble \label{EM}}

One of the most important tasks in order to connect the dark bubble model with phenomenology is including the Standard Model of particle physics coupled to the induced braneworld gravity. As a first step, the realization of electromagnetism has been carried out in \cite{Basile:2023tvh}. In this context the NS-NS $B$-field of type IIB string theory plays a crucial role, mediating the interplay between the brane and the bulk. Worldvolume fields backreact on the ambient universe in which the bubble expands, which in turn affects the energy-momentum distribution and the effective gravity induced on the brane.

The presence of a four-dimensional electromagnetic field is a well-known feature of D-branes. In particular, the Dirac-Born-Infeld action for a D3-brane reads
\begin{equation}
    S_{D_3}=-T_3 \int d^4x\sqrt{-\det(g_4+\tau \mathcal{F})} \, ,
\end{equation}
where, following \cite{Basile:2023tvh}, we define $\tau=2\pi\alpha'$ with $\alpha'\equiv l_s^2$ the (reciprocal) string tension. The D3-brane is $T_3=\frac{1}{(2\pi)^3\alpha'^2g_s}$ and the combination $\tau\mathcal{F}=\tau F+B$ contains the Maxwell curvature $F$ together with the $B$-field. At low curvatures with respect to the string scale,
\begin{align}
    S_\text{DBI} &\sim -\int d^4x\sqrt{-\det(g_4)}\left(T_3+\frac{1}{4g^2}\mathcal{F}_{\mu\nu}\mathcal{F}^{\mu\nu}\right), 
\end{align} \label{eq:DBI}
where $g^2\equiv2\pi g_s$ is the gauge coupling. Notice that the conventions are such that $\tau\mathcal{F}$ is dimensionless.

However, this is not enough: since the induced gravity on the braneworld is not described by a four-dimensional \ac{EFT}, its coupling of $F$ in the dark bubble model is not \emph{a priori} guaranteed to be the minimal coupling of Einstein-Maxwell theory. The bulk equations we need to solve arise from the bulk-brane action
\begin{equation}
     S_5=\frac{1}{2\kappa_5}\int d^5x\sqrt{-\det(g_5)}\left(R-\frac{1}{12g_s}H^2 \right) + S_\text{DBI} \, .
\end{equation}
Varying this expression, it follows that the appearance of the $B$-field in the worldvolume action induces a discontinuity across the brane \cite{Basile:2023tvh}
\begin{equation}
\Delta H^{r\mu\nu}\big|_{r=a}=H^{r\mu\nu}\big|_{r=a}=\frac{\kappa_5ka}{\pi^2\alpha'}\mathcal{F}_{\mu\nu}\big|_{r=a} \, , \label{eq:Delta}
\end{equation}
which means that the electromagnetic waves on the brane source the $H$-field in the bulk. 
The Israel junction conditions then yield
\begin{align}
    \left(\frac{\dot a}{a}\right)^2\delta^a_b & =\frac{\Lambda_4}{3}\delta^a_b+\epsilon^2\frac{\kappa_4}{6\kappa_5k^3a^4}(3\delta^a_0\delta^0_b-\delta^a_i\delta^i_b)\left(q_2+\log\left[-\xi\frac{k_+}{H}\right]\right) \\ \nonumber
    &+\frac{\kappa_4}{3}\left(\mathcal{F}^{ac}\mathcal{F}_{bc}-\frac{1}{4}\delta^a_b\mathcal{F}_{ij}\mathcal{F}^{ij}\right),
\end{align}
where the constant $q_2$ can be fixed by matching with the five-dimensional backreaction of the $B$-field \cite{Basile:2023tvh}. This result shows that $F$ couples to the induced gravity according to standard Einstein-Maxwell theory at large distances.

\subsubsection*{Fiat lux including Ramond-Ramond fluxes} \label{sec:EM_RR}

The above analysis highlights the importance of the $B$-field in the type IIB construction of \cite{Danielsson:2022lsl}. However, the type 0'B construction at the core of this work does not contain the $B$-field, due to the orientifold projection. Does this mean that the Einstein-Maxwell coupling does not arise?

\begin{figure}[ht!]
    \centering \includegraphics[width=0.8\textwidth]{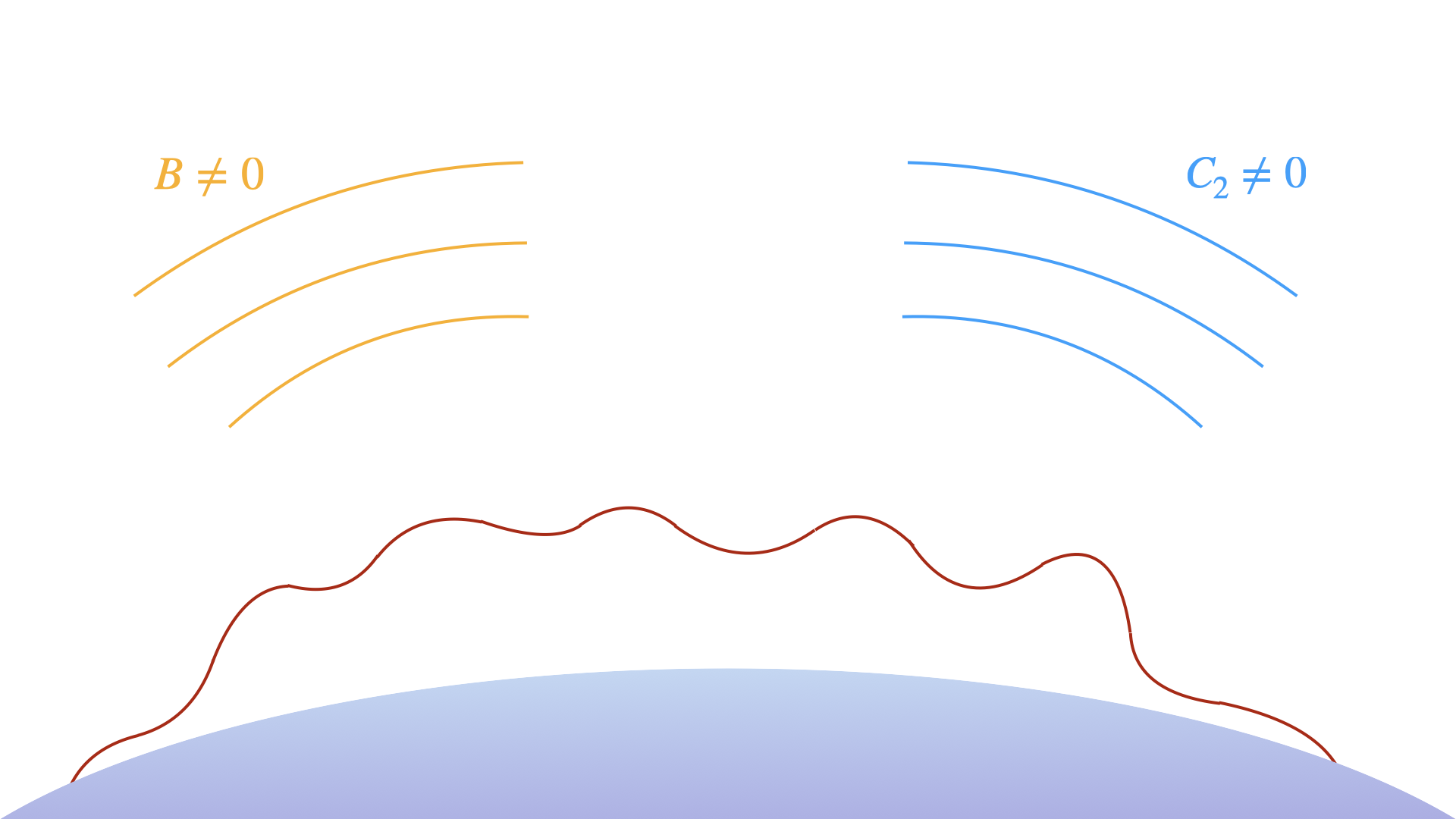}
    \caption{Representation of the mechanism induced by the electromagnetic field. Electromagnetic waves on the brane source both the $B$ and the $C_2$-field in the bulk.}
    \label{fig:fields}
\end{figure}

In order to address this issue, let us recall that bulk R-R potentials $C_n$ also couple to the worldvolume gauge field via the Chern-Simons term. For a D$p$-brane sweeping a worldvolume $W$, we have \cite{Morales:1998ux}
\begin{equation}
S_\text{CS}=\mu_p\bigoplus_n \int_{W}C_n\wedge\sqrt{\frac{\hat A(TW)}{\hat A(NW)}}\wedge e^{\tau \mathcal{F}} \, ,
\end{equation}
where the minimal charge $\mu_p=\frac{1}{(2\pi)^p l_s^{p+1}}$. Expanding the Chern character in the above expression, one finds
\begin{equation}
    S_\text{D3}=-T_3\int_{W}d^{4}x\sqrt{-\det(g_{\mu\nu}+\tau \mathcal{F}_{\mu\nu})}+\mu_3\int_{W}(C_4+\tau \mathcal{F}\wedge C_2) \, .
\end{equation}
The total bulk-brane action now takes the form \cite{Johnson:2000ch}
\begin{equation}
    S_5=\frac{1}{2\kappa_5}\int d^5x\sqrt{-g_5}\left(R-\frac{1}{12g_s}H^2-\frac{1}{12g_s}F_{3}^2 -\frac{1}{480g_s}F_{5}^2\right)+ S_\text{D3} \, .
\end{equation}
Expanding the $\mathcal{F} \wedge C_2$ as
\begin{equation}
    \tau \mathcal{F}\wedge C_2=(\tau F+B)\wedge C_2=\frac{1}{4}(\tau F_{\mu\nu}+B_{\mu\nu})C_{\rho\sigma}dx^\mu\wedge dx^\nu\wedge dx^\rho\wedge dx^\sigma \, ,
\end{equation}
the equations of motion for $B$ imply
\begin{align}
\partial_rH^{r\mu\nu}&=\frac{\kappa_5kr}{4\pi^3\alpha'^2}\left[\epsilon^{\mu\nu\rho\sigma}C_{\rho\sigma}+2\tau\mathcal{F}^{\mu\nu}\right]\delta(r-a[\eta]) \, , \nonumber\\
&= \frac{\kappa_5kr}{4\pi^3\alpha'^2}\left[\tilde C^{\mu\nu}+2\tau\mathcal{F}^{\mu\nu}\right]\delta(r-a[\eta]) \, .
\end{align}
Similarly, for $C_2$ we obtain
\begin{align}
\partial_rF^{r\rho\sigma}&=\frac{\kappa_5kr}{4\pi^2\alpha'}\mathcal{F}_{\mu\nu}\epsilon^{\mu\nu\rho\sigma}\delta(r-a[\eta]) \, ,  \nonumber \\
&= \frac{\kappa_5kr}{4\pi^2\alpha'}\mathcal{\tilde F}^{\rho\sigma}\delta(r-a[\eta]) \, .
\end{align}
Now, assuming $B$ and $C_2$ vanish on the brane, the above expressions integrate to the discontinuities
\begin{equation}
\Delta H^{r\mu\nu}=\frac{\kappa_5ka}{\pi^2\alpha'}\mathcal{F}^{\mu\nu} \, , \qquad \Delta F^{r\rho\sigma}=\frac{\kappa_5ka}{4\pi^2\alpha'}{\tilde{\mathcal{F}}}^{\rho\sigma} \, . \label{eq:dual}
\end{equation}
Therefore, $H = dB$ in the bulk is sourced by the electromagnetic field-strength on the brane, while $F_3 = dC_2$ in the bulk is sourced by its electromagnetic dual, as depicted in \cref{fig:fields}.

The upshot is that the invariance of the energy-momentum tensor under electromagnetic duality guarantees that the R-R field $C_2$ also induces an Einstein-Maxwell coupling in the braneworld. The validity of this mechanism with the $C_2$ field is of crucial importance for our dark bubble construction in \cref{sec:0'B_dS}, since the $B$-field is absent.

As a closing remark, we would like to highlight that a non-Abelian extension of the above mechanism appears unlikely to produce minimal couplings. Indeed, a discontinuity in the bulk fields sourced by a Yang-Mills field would require a gauge-invariant combination which is linear in the curvature. However, at least without including additional charged fields, there is no such combination since $\tr F = 0$ for the $SU(N)$ sector of the worldvolume gauge theory. The first non-vanishing contributions would arise from higher-order contractions, such as higher Chern classes); these would correspond to a non-minimal coupling to the induced gravity on the braneworld. It would be interesting to explore the phenomenological viability of this scenario, in particular the role played by confinement and the Higgs mechanism in this context\footnote{We thank U. Danielsson for correspondence on these matters.}.

\subsection{Dark bubbles and the swampland}\label{sec:swampland}

We conclude this section with some general remarks on the dark bubble scenario in relation to the swampland program. As we have discussed, the viability of this construction relies heavily on higher-dimensional physics in a way that is not encoded by a standard \ac{EFT}. On the contrary, both gravity and matter are delocalized. This means that, while swampland conditions are expected to apply to the bulk theory, they need not apply to the dark bubble itself. However, the interplay between the two entails that swampland conditions do play a role in the dark bubble, albeit somewhat indirectly. Indeed, the \ac{dS} conjecture \cite{obied2018sitterspaceswampland} (even in its milder formulation) was a major motivation for other attempts to realize a (quasi-)\ac{dS} cosmology beyond the standard paradigm of compactifications. Regarding the related trans-Planckian censorship conjecture of \cite{Bedroya:2019snp}, we observe that by the time the bubble wall reaches the \ac{AdS} boundary, roughly a Hubble time passes in the four-dimensional cosmology \cite{Banerjee:2019fzz}. This is consistent with the idea that a (quasi-)\ac{dS} phase last up to a Hubble time or so \cite{Bedroya:2019snp}, or more generally that its lifetime be bounded by some power of the Hubble time \cite{Dvali:2017eba, Dvali:2018jhn}.

Moreover, as we discussed in \cref{sec:probe}, the weak gravity conjecture turned out to be essential for the decay of the initial non-supersymmetric AdS vacuum, in the spirit of \cite{Ooguri:2016pdq}, and the realization of the dark bubble scenario. From a bottom-up perspective, the kinematic existence of the nucleated branes breaking the higher-form symmetry associated to the flux is consistent with the completeness of charge spectra, the absence of global symmetries and, relatedly, the triviality of cobordism classes \cite{McNamara:2019rup}.

These connections do not seem to extend to the swampland distance conjecture of \cite{Ooguri:2006in}, at least in its original formulation, since the moduli corresponding to the five internal dimensions beyond $AdS_5$ are stabilized. A discrete formulation in terms of flux quanta plays a role in the bulk \cite{Lust:2019zwm, Basile:2020mpt, Stout:2021ubb, Basile:2022zee, Stout:2022phm, Velazquez:2022eco, DeBiasio:2022zuh, Li:2023gtt, Basile:2023rvm, Palti:2024voy, Mohseni:2024njl}, but there does not seem to be any particularly strong connection with four-dimensional physics on the brane.

These general considerations are not restricted to the dark bubble, which is a rather complicated dynamical scenario without \ac{SUSY}. It is thus natural to begin a systematic investigation of the relation between swampland conditions and braneworld-like settings, as recently discussed in \cite{Anastasi:2025puv}.

\section{Quasi-de Sitter in 0'B string theory} \label{sec:0'B_dS}

Having presented the top-down framework and the bottom-up description of the dark bubble, we are now ready to present a construction using D3-branes in the type 0'B Sagnotti model. As explained in the preceding sections, among the simplest models discussed in \cref{sec:non-susy_strings} it provides the only option to realize a scale-separated braneworld. As we shall see, this uniqueness also brings along rather peculiar predictions for cosmology, including a dynamical dark energy.

\subsection{D3-branes in 0'B string theory} \label{sec:D3_0'B}

For D3-branes in the type 0'B model, the relevant near-horizon geometry was studied in~\cite{Angelantonj:1999qg, Angelantonj:2000kh,Dudas:2000sn}. More specifically, the model of~\cite{Angelantonj:1999qg, Angelantonj:2000kh} involves an O3-plane, but its contribution is sub-leading for large fluxes. Therefore we neglect the global distinction between $\mathbb{RP}^5$ and the $S^5$ which would replace it without the orientifold. In~\cite{Angelantonj:2000kh} the authors found non-homogeneous deviations from $\ads_5 \times S^5$ which are suppressed, but not uniformly so, in the large-flux limit\footnote{Similar results in tachyonic type $0$ strings were obtained in~\cite{Klebanov:1998yya}.}. We follow the treatment in \cite{Basile_2021}. The string-frame metric~\cite{Angelantonj:2000kh} reads
\begin{eqaed}\label{eq:d3-branes_metric}
    ds^2 = R^2(U) \, \frac{dU^2}{U^2} + \frac{\alpha'^2 \, U^2}{R^2(U)} \, dx^2_{1,3} + \widetilde{R}^2(U) \, d\Omega_5^2 \, ,
\end{eqaed}
where the $\ads_5$ and $S^5$ curvature radii $R(U) \, , \widetilde{R}(U)$ and the dilaton $\phi(U)$ acquire a dependence on the energy scale $U$. For large $N$, one finds
\begin{align}\label{eq:d3-branes_correction}
    \frac{R^2(U)}{R^2_{\infty}} & \sim 1 - \frac{3}{16} \, g_s \, \alpha' T \, \log \left(\frac{U}{u_0}\right) \, , \\
    \frac{\widetilde{R}^2(U)}{R^2_{\infty}} & \sim 1 - \frac{3}{16 \sqrt[4]{8}} \, g_s^2 \, N \, \alpha' T \, \log \left(\frac{U}{u_0}\right) \, , \\
    \frac{1}{N} \, e^{-\phi} & \sim \frac{1}{g_s \, N} + \frac{3}{8 \sqrt[4]{8}} \, g_s \, \alpha' T \, \log \left(\frac{U}{u_0}\right) \, .
\end{align}
Here $u_0$ is a reference scale, $R^2_\infty = \sqrt{4\pi g_s N} \alpha'$ is the supersymmetric value of both radii. The validity of the EFT description requires $g_s N \gg 1$~\cite{Maldacena:1997re}, while \cref{eq:d3-branes_correction} requires $g_s^2 \, N \ll 1$. Thus, the regime of validity of \cref{eq:d3-branes_correction} is $g_s^2 N \ll 1 \ll g_s N$.

The five-form R-R field strength $F_5$ in the background is self-dual, closed\footnote{Since the orientifold projection removes the $B$-field, no additional terms appear in the Bianchi identity.}, and reads
\begin{align}\label{eq:d3_f5}
    F_5 & = \left(1 + \star\right) f_5 \, N \, \text{vol}_{S^5} \\
    & = f_5 \, N \, \text{vol}_{S^5} + \frac{f_5 \, N}{\widetilde{R}(U)^5} \left( \frac{\alpha' \, U}{R(U)}\right)^3 \, d(\alpha' U) \wedge d^4 x
\end{align}
with $\text{vol}_{S^5}$ the volume form of the unit $S^5$. The flux quantization condition
\begin{eqaed}\label{eq:f_5_flux_quantization}
    \frac{1}{2\kappa_{10}^2} \int_{S^5} F_5 = \mu_3 \, N \, ,
\end{eqaed}
where $\mu_3$ is the charge of a single D3-brane, then fixes $f_5$, and therefore the $C_4$ potential takes the form
\begin{eqaed}
    C_4 = c_4(U) \, d^4 x + \dots
\end{eqaed}
where
\begin{eqaed}
    \frac{c_4'(U)}{\alpha'} = \frac{f_5 \, N}{\widetilde{R}(U)^5} \left( \frac{\alpha' \, U}{R(U)}\right)^3 \, .
\end{eqaed}
As a result, the probe potential
\begin{eqaed}\label{eq:d3_probe_potential}
    V_{\text{probe}}^{\text{D}3}(U) = N_3 T_3 \left( \frac{\alpha' \, U}{R(U)} \right)^4 e^{-\phi(U)} - N_3 \mu_3 \, c_4(U)
\end{eqaed}
simplifies to \cite{Basile_2021}
\begin{eqaed}\label{eq:d3_probe_sub-leading}
    V_{\text{sub-leading}}^{\text{D}3}(U) \propto U^4 \left[5 - 4 \, \log \left(\frac{U}{u_0}\right) \right] \, .
\end{eqaed}
Including numerical factors, the effective Lagrangian for the dynamics of a (rigid) probe is
\begin{align}\label{eq:eff_dark bubbleI_action}
    \mathcal{L}_\text{NLO} & = \frac{N_3 N U^4}{2\pi^2 \lambda^2} \bigg[- \left(1 + \frac{3 \lambda^2 \alpha' T}{128 \sqrt[4]{8} \pi^2 N} \, \log\frac{U}{u_0}\right) \sqrt{1 - \frac{\lambda \dot{U}^2}{U^4}} \nonumber \\
    & + 1 - \frac{15 \lambda^2 \alpha' T}{2048 \sqrt[4]{8} \pi^2 N} \left(1 - 4 \, \log\frac{U}{u_0} \right) \bigg] \, ,
\end{align}
where the 't Hooft coupling $\lambda \equiv 4\pi g_s N$ and, in the 0'B model, $\alpha' T = 8/\pi^2$ \cite{Sagnotti:1995ga, Sagnotti:1996qj, Dudas:2000sn}.

\begin{figure}[ht!]
    \centering \includegraphics[width=0.7\textwidth]{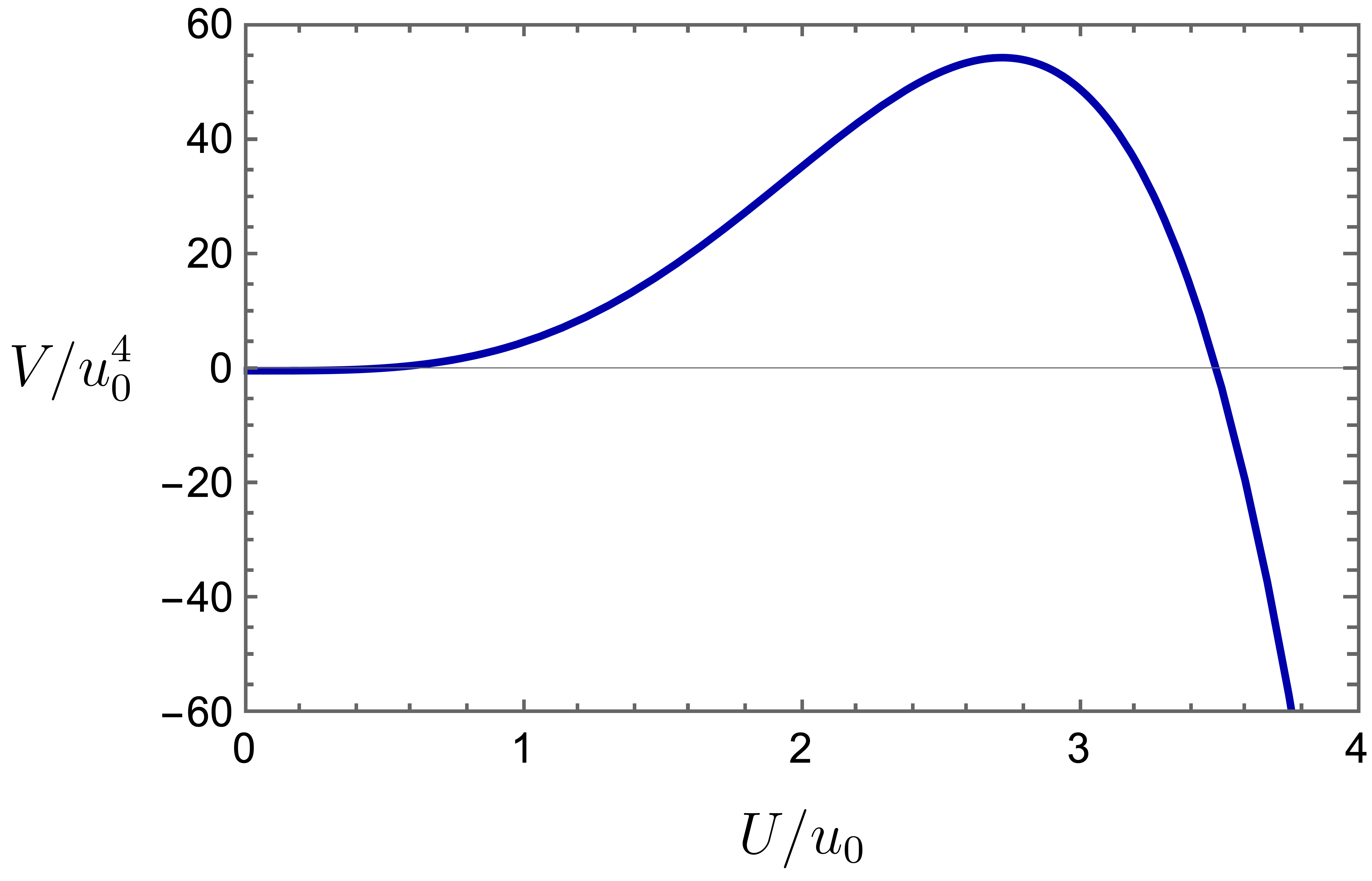}
    \caption{Normalized probe potential in \cref{eq:d3_probe_sub-leading} in units of the reference scale $u_0$.}
    \label{fig:probe}
\end{figure}

As mentioned above, the resulting geometry is not completely under control because of the non-uniform character of corrections to the AdS vacuum, but it is in principle amenable to numerical investigation via the Toda-like formalism of~\cite{Dudas:2000sn}. A first estimate of the higher-derivative corrections for large fluxes comes from on-shell ratios of the form $\frac{F_5^{2m}}{F_5^2} \approx N^{-\frac{m-1}{2}} \ll 1$ in string units, for values of $u$ within the controlled large-$N$ regime. Since $dF_5$ vanishes on shell, we do not expect derivative terms to affect the solution to leading order either.
\subsection{Realizing the dark bubble scenario} \label{sec:0'B_dark bubblec}

The worldvolume action can also be used to obtain Friedmann equations for the braneworld model, casting the induced metric on the brane in a cosmological guise. For the Poincaré slicing we employed in \cref{eq:d3-branes_metric}, the appropriate form is $ds^2_\text{brane} = - d\tau^2 + a(\tau)^2 d\mathbf{x}^2$ with flat spacelike slices.

From the equation of motion of the NLO brane Lagrangian one can recover the Friedmann equation with the precise numerical factors. The pullback of the ten-dimensional metric gives a four-dimensional cosmology with scale factor
\begin{eqaed}
\label{eq:cosmfactorrel}
    a(t) = \frac{\alpha' U(t)}{R(U(t))}\,,
\end{eqaed}
which we can recast in terms of the cosmological time defined by $\left(\frac{d\tau}{dt}\right)^2 = - \, \frac{R(U)^2}{U^2} \dot{U}^2 + \frac{\alpha'^2 U^2}{R(U)^2}$. The Lagrangian in \cref{eq:eff_dark bubbleI_action} yields the on-shell conserved Hamiltonian
\begin{eqaed}\label{eq:NLO_hamiltonian}
    \mathcal{H}=\dot{U}\dfrac{\partial\mathcal{L}}{\partial \dot{U}}-\mathcal{L} \sim \frac{N_3 N U^4}{2 \pi^2 \lambda^2} \left[\frac{1}{\sqrt{1 - \, \frac{\lambda \dot{U}^2}{U^4}}} - \, 1 + \frac{3 \, \lambda^2 \, \alpha' T}{2048 \sqrt[4]{8} \pi^2 N}\left(5 - 4 \, \log \frac{U}{u_0} \right)\right],
\end{eqaed}
once again to the first subleading order. For convenience, we define
\begin{align}\label{eq:eps_precise}
    \epsilon & \equiv \frac{3 \, \lambda^2 \, \alpha' T}{256 \sqrt[4]{8} \pi^2 N} = \frac{3 \, \alpha' T}{16 \sqrt[4]{8}} \, g_s^2 N \overset{\text{0'B}}{=} \frac{3}{2^{\frac{7}{4}} \pi^2} \, g_s^2 N \approx 0.09 \, g_s^2 N \, ,
\end{align}
such that the Hamiltonian takes the simple form
\begin{eqaed}\label{eq:NLO_hamiltonian_eps}
    \mathcal{H} \sim \frac{N_3 N U^4}{2 \pi^2 \lambda^2} \left[\frac{1}{\sqrt{1 - \, \frac{\lambda \dot{U}^2}{U^4}}} - \, 1 + \frac{\epsilon}{2} \left(\frac{5}{4} - \log \frac{U}{u_0} \right)\right] .
\end{eqaed}
By conservation, the term in square brackets has to vanish at late times, since the prefactor $U^4$ blows up as the brane expands. This implies
\begin{eqaed}
    a \sim \frac{\alpha'U}{R_\infty} \, , \qquad\left(\dfrac{1}{a}\dfrac{da}{d \tau}\right)^2\equiv H^2 \sim \frac{\lambda \dot{U}^2}{R_\infty^2 U^4} \sim \epsilon R_\infty^{-2} \log\left( \frac{a}{a_*} \right) \, ,
\end{eqaed}
where we have used \cref{eq:cosmfactorrel}. Therefore, we obtain
\begin{eqaed}\label{eq:friedmann_eqs}
    H^2 = \epsilon M^2 \log\left( \frac{a}{a_*} \right) \, ,
\end{eqaed}
where $M \equiv R_\infty^{-1}$ denotes the scale of higher-dimensional physics\footnote{If there were an EFT description on the braneworld, $M$ would be its cutoff.}. The arbitrary reference scale $a_*$ disappears by fixing the initial time so that $a(0) = a_0$ is the scale factor today, and similarly $H(0) = H_0$. The resulting (today-accelerating branch of the) solution takes the form
\begin{eqaed}\label{eq:scale_factor}
    a(\tau) = a_0 \, \exp(H_0 \, \tau + \frac{\epsilon M^2 \tau^2}{4}) \, ,
\end{eqaed}
whereas the branch that decelerates at present time would have a minus sign in front of the linear term in $\tau$.

\Cref{eq:friedmann_eqs} has a remarkable feature: the logarithmic running of dynamical dark energy arises from the running of the classically dimensionless gauge coupling on the worldvolume. This explains why D3-branes are the only reliable option for this kind of highly non-supersymmetric dark bubble realization from a holographic perspective \cite{Angelantonj:1999qg, Angelantonj:2000kh}. This consideration follows from general grounds, and thus the logarithmic running of dark energy is expected to arise in any realization of this type. By contrast, the corrections involved in the type IIB construction of \cite{Danielsson:2022lsl} are of higher-derivative type.

\subsection{On effective realizations via scalar fields} \label{sec:effective_scalars}

The form of the scale factor in \cref{eq:scale_factor} indicates a quasi-\ac{dS} expansion. Although, as we have discussed, there is no \ac{EFT} realization of this dark bubble construction, it is natural to ask whether the dynamical dark energy in \cref{eq:friedmann_eqs} affords some classical four-dimensional description as for electromagnetic fields \cite{Basile:2023tvh}. To this end, we would like to understand whether it is possible to reproduce this behavior via a homogeneous scalar field $\phi$ subject to a potential $V(\phi)$.

Consider a scalar field $\phi(\textbf{x},t)$. Using the FLRW metric with a flat slicing,
\begin{eqaed}
    ds^2=-c^2dt^2+a^2(t)d\textbf{x}^2 \, ,
\end{eqaed}
the action for homogeneous configurations $\phi(\textbf{x},t)=\phi(t)$ evaluates to
\begin{eqaed}
    S=\int d^4x a^3(t)\left[\frac{1}{2}\partial_\mu\phi\partial^\mu\phi-V(\phi)\right] = \int d^4x a^3(t)\left[\frac{1}{2}\dot\phi^2-V(\phi)\right] .
\end{eqaed}
The resulting equation of motion
\begin{eqaed}
    \ddot\phi+3H \dot\phi+ \frac{\partial V}{\partial \phi}=0
\end{eqaed}
is accompanied by the Friedmann equations
\begin{eqaed}
\label{eq:Friedmanneqs1}
   \left(\dfrac{\dot{a}}{a}\right)^2\equiv H^2=\frac{1}{3 M_\text{Pl}^2}\rho_\phi=\frac{\frac{1}{2}\dot\phi^2+V(\phi)}{3 M_\text{Pl}^2}
\end{eqaed}
and
\begin{eqaed}
\label{eq:Friedmanneqs2}
\dot{H}=-\frac{1}{2 M_\text{Pl}^2}\left(\rho_\phi+P_\phi\right)=-\frac{\dot\phi^2}{2 M_\text{Pl}^2} \, .
\end{eqaed}
Here we have defined the pressure and density of the homogeneous scalar field. From \cref{eq:Friedmanneqs1} and \cref{eq:scale_factor} one can notice an immediate contradiction. For the 0'B dark bubble $\dot{H}$ is strictly positive, while $-\dot\phi^2$ is strictly negative. As such, it is impossible to describe the 0'B dark bubble as an effective theory of single field inflation.

\subsubsection*{No-go for scalar potentials}

One can attempt to circumvent the above single-field result seeking an effective description in terms of a general multifield scalar theory. Consider the Lagrangian density
\begin{eqaed}
    \mathcal{L}=\frac{1}{2}g_{ij}(\phi)\partial_\mu\phi^i\partial^\mu\phi^j-V(\phi) \, ,
\end{eqaed}
where $g_{ij}$ is a metric in field space and the spacetime metric is encoded in the partial derivatives of $\phi$. The Euler-Lagrange equations yield
\begin{align}
\Box\phi^k &+\Gamma_{ij}^k\partial_\mu\phi^i\partial^\mu\phi^j=-g^{kj}(\phi)\frac{\partial V}{\partial \phi^j} \, , \label{eq:box}\\
    H^2 &= \frac{1}{3 M_p^2}\left(V(\phi)+\frac{1}{2}g_{ij}
(\phi)\dot\phi^i\dot\phi^j\right) \, , \label{eq:ok}
\end{align}
where the field-space Christoffel symbols are defined as $\Gamma_{ij}^k=\frac{1}{2}g^{kl}(\partial_jg_{il}+\partial_ig_{jl}-\partial_lg_{ij})$.
Since we are interested only in spatially homogeneous solutions, $\phi^i(\textbf{x},t)=\phi^i(t)$, \cref{eq:box} becomes
\begin{eqaed}
 \label{eq:gammagen}
\ddot\phi^k+3H\dot\phi^k+\Gamma_{ij}^k\dot\phi^i\dot\phi^j=-g^{kj}(\phi)\frac{\partial V}{\partial \phi^j} \, .
\end{eqaed}
On the other hand, taking the time derivative of \eqref{eq:ok}, we get
\begin{eqaed}
\label{eq:hdotgen}
    \frac{\partial V}{\partial\phi_k}\dot\phi^k+g_{ij}\ddot\phi^i\dot\phi^j=3 M_p^2(2H\dot H)-g_{li}\Gamma_{kj}^l\dot\phi^i\dot\phi^j\dot\phi^k \, .
\end{eqaed}
Finally, substituting \cref{eq:gammagen} in \cref{eq:hdotgen}, we find that the terms depending on the connection vanish. As a result
\begin{eqaed}
    2 M_p^2\dot H+3g_{ij}\dot\phi^i\dot\phi^j=0\, ,
\end{eqaed}
and we recover \cref{eq:Friedmanneqs2} in this more general setting. We can thus conclude that no effective scalar potential is able to reproduce the time evolution of the braneworld \eqref{eq:scale_factor}. Equivalently no effective scalar potential can be used to construct a super-exponential cosmological acceleration phase.

\subsubsection*{Invoking phantom scalars}

The above analysis shows that for models of scalar-driven accelerated expansion, the first derivative of the Hubble radius must necessarily decrease. This can also be directly seen from the equation of state
\begin{eqaed}
    w=\frac{P_\phi}{\rho_\phi}=\dfrac{\frac{1}{2}\dot\phi^2-V(\phi)}{\frac{1}{2}\dot\phi^2+V(\phi)}\geq -1 \, ,
\end{eqaed}
where \ac{dS} is the limiting case for which $\dot\phi=0$. We recall how different cosmological solutions depend on the equation of state according to

\begin{itemize}
    \item $w>-1:$ \qquad $a\simeq \tau^{\frac{2}{3(w+1)}}$ \, ,
    \item $w=-1:$ \qquad $a\simeq e^{H_0 \tau}$ \, ,
    \item $w<-1:$ \qquad $a\simeq (\tau_s-\tau)^{\frac{2}{3(w+1)}}$ \, ,
\end{itemize}
where $\tau_s$ denotes the time of the singularity, which arises in models with $w<-1$. In such a universe, the comoving particle horizon shrinks to zero size in finite time, and all objects in the universe would be causally disconnected. This is usually referred to as Big Rip. From the Friedmann equations (here we use $M_\text{Pl}^{-2}=8\pi G$)
\begin{eqaed}
    H^2=\dfrac{\rho}{3 M_\text{Pl}^2},\,\quad\, \dot{H}=-\dfrac{\rho}{2 M_\text{Pl}^2}(1+w)
\end{eqaed} 
we can derive the  relation
\begin{eqaed}
   w=-1-\dfrac{2}{3}\dfrac{\dot{H}}{H^2} \, ,
\end{eqaed}
from which we compute the equation of state from \cref{eq:scale_factor}. We obtain
\begin{eqaed}
\label{eq:0beqmotion}
    w_\text{0'B}=-1-\frac{\epsilon M^2}{3 (H_0+\dfrac{\epsilon M^2}{2} \tau)^2} \, .
\end{eqaed}
Therefore, in the 0'B dark bubble the equation of state always satisfies $w<-1$, approaching pure \ac{dS} ($w=-1$) at late times $\tau\to\infty$. In fact, the logarithmic dependence on the scale factor in \cref{eq:friedmann_eqs} encodes that the spacetime singularity sits at infinite cosmic time, and furthermore that the slow-roll parameter vanishes there. This corresponds to a specific scenario of Big Rip often dubbed Little Sibling of Big Rip (LSBR). This behaviour corresponds to an abrupt event rather than a future spacetime singularity. At this event, the Hubble rate and the scale factor blow up but the cosmic derivative of the Hubble rate does not. Consequently, this abrupt event takes place at an infinite cosmic time where the scalar curvature diverges.

Although we have just proved a no-go result for regular scalar fields, a simple realization of such a scenario involves a phantom field, namely a canonically normalized scalar field with a potential, but with the ``wrong'' sign in the kinetic energy. In this sense, the phantom scalar field can be viewed as a particular case of the K-essence models with $K(X)=-X$. Its action is
\begin{eqaed}
    S=\int d^4x a^3(t)\left[-\frac{1}{2}\partial_\mu\phi\partial^\mu\phi-V(\phi)\right] .
\end{eqaed}
As described in \cite{Bouhmadi-Lopez:2018lly}, this can also be described by a four-dimensional theory including only gravity and a three-form field with a potential. Such a theory does not lead to a loss of unitarity, but can, in some regimes, violate the Wheeler-DeWitt equation. For cosmological solutions the above action simplifies to
\begin{eqaed}
    S=\int d^4x a^3(t)\left[-\frac{1}{2}\dot\phi^2-V(\phi)\right],
\end{eqaed}
leading to the equation of state
\begin{eqaed}
    w=\frac{P_\phi}{\rho_\phi}=\dfrac{-\frac{1}{2}\dot\phi^2-V(\phi)}{-\frac{1}{2}\dot\phi^2+V(\phi)} \, .
\end{eqaed}
We can see that for dominant potential energy, $V(\phi)>\dot\phi^2/2$, the equation of state always satisfies $w<-1$. For a Hubble scale $H=H_0+\epsilon M^2\frac{\tau}{2}$, which arises from \cref{eq:scale_factor}, the Friedmann equations for the phantom scalar lead to
\begin{eqaed}
   V(\phi)=3 M_\text{Pl}^2\left(H_0+\epsilon M^2\frac{\tau}{2}\right)^2+\frac{1}{2}\dot\phi^2 ,\,\qquad\,\dot\phi=\sqrt{\dfrac{\epsilon M^2 M_\text{Pl}^2}{4}} \, .
\end{eqaed}
Appropriately choosing the integration constant for $\phi$ such that
\begin{eqaed}
  \phi=\frac{2H_0 M_\text{Pl}}{\sqrt{\epsilon M^2}} + \sqrt{\dfrac{\epsilon M^2 M_\text{Pl}^2}{4}}\tau \, ,
\end{eqaed}
we can finally recast our results as a theory of cosmic acceleration driven by a phantom scalar with a quadratic potential
\begin{eqaed}
\label{eq:V_quadratic}
    V(\phi)=\epsilon \dfrac{M^2}{2} M_\text{Pl}^2+ \epsilon\dfrac{3 M^2}{4} \phi^2 \, ,
\end{eqaed}
as previously noted in \cite{Albarran:2016mdu}. Now, it is clear that a solution to \cref{eq:Friedmanneqs2} exists precisely for the phantom scalar considered. Within slow-roll approximation, the equation of motion reads
\begin{eqaed}
    \ddot\phi+3H\dot\phi-V'\simeq 3H\dot\phi-V'=0 \, .
\end{eqaed}
Using this, we can write the slow-roll parameters for the phantom scalar according to
\begin{eqaed}
    \varepsilon=-\dfrac{\dot{H}}{H^2}\simeq-\dfrac{1}{2}\dfrac{\dot\phi^2}{H^2}\simeq\varepsilon_V=-\dfrac{1}{2}\left(\dfrac{V'}{V}\right)^2 \, , \label{eq:epsilon}
\end{eqaed}
\begin{eqaed}
    \eta=-\dfrac{\dot{\varepsilon}}{H\varepsilon}\simeq -2\left(\dfrac{V'}{V}\right)^2+2\left(\dfrac{V''}{V}\right)\simeq 4\varepsilon_V-2\eta_V \, , \label{eq:eta}
\end{eqaed}
where for simplicity we take $M_\text{Pl}=1$.

\section{Phenomenological aspects} \label{sec:pheno}

Having examined the dark bubble realization in the type 0'B Sagnotti model, as well as an effective realization of its dynamical dark energy, we now turn to phenomenological bounds. Our aim is to derive the analogous scalings to \cref{eq:iib_scalings} in our setting and to study the viability of this model for early- and late-time cosmology.

\subsection{Scales and bounds for particle physics} \label{sec:scales_bounds}

In order to derive phenomenological bounds and visualize them, in this section we replace parametric bounds with (order-of-magnitude) estimates. Namely, formulae such as $x \ll 1$ are replaced by $x \lesssim 1$ for the sake of this section. The validity of the ten-dimensional solution of \cref{eq:d3-branes_metric} requires
\begin{equation}
g_s^2 N \log\frac{U}{u_0} \lesssim 1
\end{equation}
which, to leading order, translates into
\begin{equation}
\epsilon  \log \frac{a}{a_*} \lesssim 1 \, .
\end{equation}
Since the scale factor $a_*$ is fixed by the initial conditions as
\begin{equation}
a_* = a_0 \exp\left(- \frac{H_0^2}{\epsilon M^2}\right),
\end{equation}
one finds that the cosmological time is bounded as
\begin{equation}
\tau \lesssim \frac{2}{\epsilon M} \left(1 - \frac{H_0}{M}\right)
\end{equation}
to leading order in $\epsilon$. Consequently, the Hubble rate in this regime satisfies
\begin{equation}
H \lesssim M,
\end{equation}
ensuring the reliability of an effective four-dimensional description. This serves as a reassuring, albeit expected, cross-check.

Requiring that the maximal cosmological time for EFT validity exceeds the age of the universe,
\begin{equation}
\frac{1}{H_0} \lesssim \frac{2}{\epsilon M} \left(1 - \frac{H_0}{M}\right),
\end{equation}
leads to the bound
\begin{equation}
H_0 \lesssim M \lesssim H_0/\epsilon \, .
\end{equation}
Notably, the upper bound on $M$ follows without assuming a specific value of $a_*$ at leading order in $\epsilon$, as expected.

It would be intriguing to explore whether the trans-Planckian censorship conjecture (TCC) provides additional constraints on the cosmological constant, though its applicability remains uncertain as we discussed in \cref{sec:swampland}. While the full geometry sourced by the brane is more complex and is expected to culminate in a dynamical cobordism \cite{Antonelli:2019nar, Buratti:2021fiv, Blumenhagen:2022mqw, Angius:2022aeq, Blumenhagen:2023abk, Angius:2023xtu, Huertas:2023syg, Calderon-Infante:2023ler, Angius:2023uqk, Angius:2024zjv} (where a strongly coupled regime emerges), the \ac{AdS} coordinate time $t$ of the entire expansion can be estimated by approximating the geometry as pure \ac{AdS}, which is
\begin{equation}
t = \frac{\pi}{2} R_\infty \, .
\end{equation}
The above bounds imply that this duration is at most the Hubble time $H_0^{-1}$, aligning with certain implications of the TCC regarding the lifetime of a quasi-\ac{dS} universe from a bulk perspective. This was already observed in \cite{Banerjee:2019fzz}.

In string units, where $M_s = \alpha'^{-1/2} = 1$, the characteristic energy scales exhibit the following dependencies on $g_s$ and $N$:
\begin{align}
M & = \frac{1}{R_\infty} \propto g_s^{-\frac{1}{4}} N^{-\frac{1}{4}}, \
M_\text{Pl} & \sim \left(\frac{R^6}{g_s^2 N}\right)^{\frac{1}{2}} \propto g_s^{-\frac{1}{4}} N^{\frac{1}{4}}, \
M_5 & \sim \left(\frac{R^5}{g_s^2}\right)^{\frac{1}{3}} \propto g_s^{-\frac{1}{4}} N^{\frac{5}{12}} \, ,
\end{align}
where $M_5$ is the five-dimensional Planck scale and $M_\text{Pl}$ denotes the four-dimensional Planck scale obtained from the dark bubble relation \cite{Banerjee:2019fzz}
\begin{equation}
M_\text{Pl}^2 = M_5^3 \delta R \sim M_5^3 R/N \, .
\end{equation}
Expressed in four-dimensional Planck units, these relations yield
\begin{align}
\frac{M}{M_\text{Pl}} \sim N^{-\frac{1}{2}} \ll 1\,,\qquad 
\frac{M_5}{M_\text{Pl}}\sim N^{\frac{1}{6}} \gg 1 \, .
\end{align}
Given that perturbative control requires $g_s N \gtrsim 1$, we find $\epsilon \gtrsim 1/N$, leading to the bounds
\begin{equation}
\frac{H_0}{M_\text{Pl}} \lesssim \frac{M}{M_\text{Pl}} \lesssim \frac{H_0}{\epsilon M_\text{Pl}} \lesssim \frac{H_0}{M_\text{Pl}}  N.
\end{equation}
Thus, in Planck units, the constraints simplify to
\begin{equation}
\Lambda^{\frac{1}{2}} \sim H_0 \lesssim M \lesssim H_0^{\frac{1}{3}} \sim \Lambda^{\frac{1}{6}},
\end{equation}
with $\Lambda$ the present day dark energy. Although we will generally consider the implications of the 0'B dark bubble for inflation, here we focus on constraints for late-time acceleration i.e. dark energy. Comparing the above result with the considerations in~\cite{Montero:2022prj}, the lower bound on $M$ is consistent with the Higuchi bound, while the upper bound in our case is $M \sim \Lambda^{\frac{1}{6}} \, M_\text{Pl} \approx 100 \text{ MeV}$ instead of $\Lambda^{\frac{1}{4}} \, M_\text{Pl} \approx 3 \text{ meV}$.

As we have discussed in \cref{sec:swampland}, swampland and/or \ac{EFT} arguments are not expected to apply to the dark bubble, so there is no contradiction. Nevertheless, the validity of these arguments for a braneworld model, as opposed to a compactification, should be assessed more carefully. For instance, the argument for the upper bound with exponent $\frac{1}{d} = \frac{1}{4}$ stems from considerations on the effective potential generated by a tower of light states, whose appearance, according to the distance conjecture(s), generally affects the bulk rather than a braneworld. The smallness of the quasi-dS dark energy in our construction is not necessarily tied to the smallness of the bulk \ac{AdS} scale, although in this case they both are. It is thus worth considering whether the \ac{AdS} version of the distance conjecture \cite{Lust:2019zwm} applies, even though it is not expected to. In particular, in the bulk (quasi-)$\ads_5$ there is a light tower of \ac{KK} states with masses $m_\text{KK} \sim R_\infty^{-1} \sim M$, and the bound of~\cite{Montero:2022prj} would then take the form $M/M_5 \lesssim (\Lambda_\ads/M_5^2)^{\frac{1}{5}} \sim (M/M_5)^\frac{2}{5}$, which is satisfied since $M/M_5 \ll 1$. On the other hand, if this tower dominates the species bound \cite{dvali2010speciesstrings}, one would naively obtain the species scale
\begin{equation}\label{eq:species_scale}
    \Lambda_\text{sp} = \frac{M_\text{pl}}{\sqrt{N(\Lambda_\text{sp})}} \sim M^{\frac{5}{7}} \lesssim \Lambda^{\frac{5}{42}} \approx 100 \text{ TeV}
\end{equation}
as a quantum gravity cutoff in the spirit of~\cite{Montero:2022prj}. However, we stress once again that the dark bubble is not expected to be described by a \emph{bona fide} \ac{EFT} in four dimensions.

It turns out that the cosmological expansion yields stronger bounds. The slow-roll parameter today is given by
\begin{equation}
\varepsilon_H = \frac{\dot H}{H_0^2} = \frac{\epsilon M^2}{2H_0^2} \sim \epsilon \frac{M^2}{\Lambda} \, ,
\end{equation}
implying the bound

\begin{equation}
\Lambda \lesssim \varepsilon_H \lesssim \Lambda^{- \frac{2}{3}} \, .
\end{equation}
In Planck units, this two-sided bound spans the very large range
\begin{equation}
10^{-122} \lesssim \varepsilon_H \lesssim 10^{81} \, .
\end{equation}
For compatibility with standard cosmological models, we impose $\varepsilon_H \lesssim 1$, leading to the constraint
\begin{equation}
\Lambda^{\frac{1}{2}} \lesssim M \lesssim \frac{H_0}{\sqrt{\epsilon}} \lesssim \frac{\Lambda^{\frac{1}{2}}}{M} \lesssim \Lambda^{\frac{1}{4}} \, ,
\end{equation}
which now does coincide with the range found in~\cite{Montero:2022prj}.

In the context of the braneworld model, the relevant scale of new physics is expected to be $M$, rather than $M_5$ or $\Lambda_\text{sp}$. Unlike decompactification scenarios, where a higher-dimensional \ac{EFT} remains valid, in this case resolving the extra dimensions at the scale $M$ would also reveal that the bulk geometry is (quasi-)\ac{AdS} rather than (quasi-)\ac{dS}, as suggested in~\cite{Montero:2022prj}. When including the slow-roll-like condition the upper bound for $M$ is the expected neutrino mass scale
\begin{eqaed}
    M \lesssim \Lambda^{\frac{1}{4}}\approx \text{meV} \approx m_\nu \, , 
\end{eqaed}
which would render the theory apparently invalid phenomenologically. To circumvent this potential issue, the bulk \ac{KK} modes would need to be weakly coupled to the braneworld.

\subsubsection*{String masses} \label{sec:string_masses}

The masses of particle endpoints of fundamental strings stretching to the boundary of $\ads_5$ are given by $m_s = M_s^2/M$ \cite{Banerjee:2019fzz, Banerjee:2020wov, Basile:2020mpt}. This leads to the scaling relation $m_s/M_\text{Pl} \sim g_s^{\frac{1}{2}}$. In passing, we remark that $g_s$ corresponds to the ten-dimensional dilaton. Given the range of validity $N^{-1} \lesssim g_s \lesssim N^{-\frac{1}{2}}$, one obtains:
\begin{eqaed}\label{eq:string_mass}
    \frac{M}{M_\text{Pl}} \sim N^{-\frac{1}{2}} \lesssim \frac{m_s}{M_\text{Pl}} \lesssim N^{-\frac{1}{4}} \sim \left(\frac{M}{M_\text{Pl}}\right)^{\frac{1}{2}} \, .
\end{eqaed}
In physical units, this corresponds to a mass window of
\begin{eqaed}
    10^{-41} \text{ GeV} \approx \Lambda^{\frac{1}{2}} \lesssim m_s \lesssim \Lambda^{\frac{1}{12}} \approx 10^9 \text{ GeV} \, ,
\end{eqaed}
without assuming a specific scaling for $M$. If one fixes it to the upper bound $M \sim \Lambda^{\frac{1}{6}}$, this range refines to
\begin{eqaed}
    100 \text{ MeV} \sim \Lambda^{\frac{1}{6}} \lesssim m_s \lesssim \Lambda^{\frac{1}{12}} \approx 10^9 \text{ GeV} \, .
\end{eqaed}
Furthermore, incorporating a slow-roll-like condition, $\varepsilon_H \lesssim 1$ places an additional constraint, refining the upper bound to
\begin{eqaed}
    m_s \lesssim \Lambda^{\frac{1}{8}} \approx 10 \text{ TeV} \, .
\end{eqaed}
Then, the string scale satisfies
\begin{eqaed}
    \Lambda^{\frac{1}{2}} \lesssim M_s \sim \sqrt{m_s M} \lesssim \Lambda^{\frac{1}{6}} \, ,
\end{eqaed}
while the five-dimensional Planck scale is given by
\begin{eqaed}
    M_5 \sim N^{\frac{1}{6}} \sim M^{-\frac{1}{3}} \lesssim \Lambda^{-\frac{1}{6}} \approx 10^{39} \text{ GeV} \, ,
\end{eqaed}
which is extremely large—a common feature in dark bubble constructions. In contrast to compactification scenarios, this does not suggest that the higher dimensional bulk gravitational effects corresponding to the ambient $AdS_5$ will be highly suppressed. At any rate, as already mentioned, \ac{KK} modes still present a threat to this sort of models, unless somehow suppressed. 

\subsubsection*{Putting the bounds together} \label{sec:bounds}

\begin{figure}
    \centering
    \begin{tabular}{c c}  
        \includegraphics[height=0.45\textwidth]{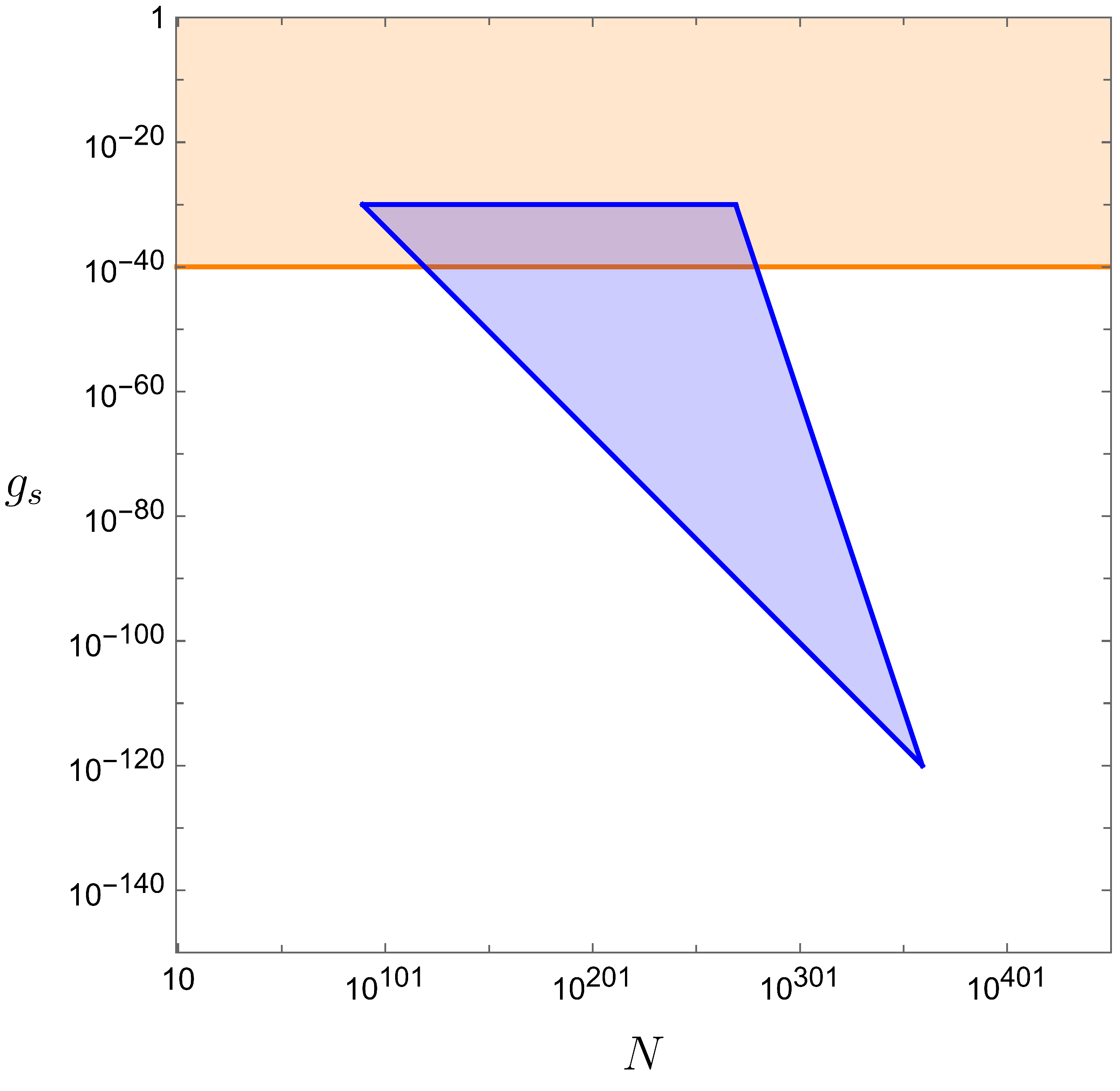} & 
        \includegraphics[height=0.45\textwidth]{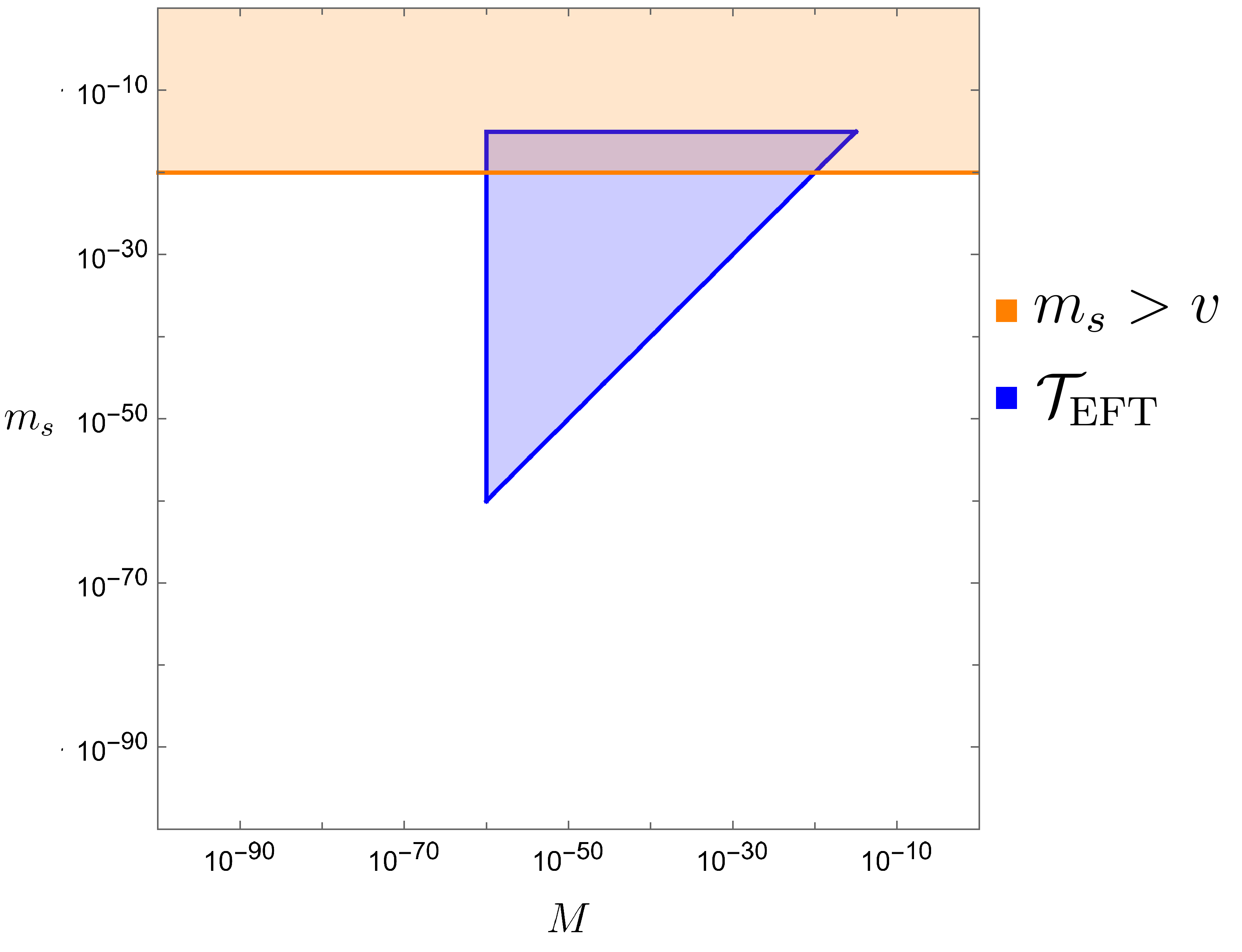} \\
    \end{tabular}
    \caption{Parameter space of the model. Left: The $(N, g_s)$-plane, showing the perturbative regime, the \ac{EFT} and slow-roll regime, and the region where Higgsing can occur at the electroweak scale. Right: The $(M, m_s)$-plane in Planck units, depicting the same regimes and constraints.}
    \label{fig:bounds}
\end{figure}

Summarizing the preceding analysis, assuming slow-roll $M$ is constrained to the range
\begin{eqaed}
    \Lambda^{\frac{1}{2}} \lesssim M \lesssim \Lambda^{\frac{1}{4}} \, ,
\end{eqaed}
while the string mass on the bubble is bounded as
\begin{eqaed}
    M<m_s\lesssim \Lambda^{\frac{1}{8}} \, ,
\end{eqaed}
where the left hand side ensures that the \ac{EFT} remains reliable. We can then write
\begin{eqaed}
    M\lesssim m_s\lesssim M^{\frac{1}{4}} \, .
\end{eqaed}
The perturbative slow-roll \ac{EFT} region forms the triangular domain
\begin{eqaed}\label{eq:pert_slow-roll_EFT_triangle}
    \mathcal{T}_\text{EFT} \equiv \left\{ M \gtrsim \Lambda^{\frac{1}{2}} \, , \quad m_s \lesssim \Lambda^{\frac{1}{8}} \, , \quad M \lesssim m_s \right\}
\end{eqaed}
in four-dimensional Planck units. 
We can also impose that branes are able to separate and generate a Higgs(-like) mechanism at a scale $v$. Since $m_s$ is the maximal mass achievable by open strings stretching between branes, this requires
\begin{eqaed}
    m_s \gtrsim v \, .
\end{eqaed}
We can express the constraints in terms of $g_s$ and $N$ by using 
\begin{eqaed}
    M\sim g_s^{-\frac{1}{4}} N^{-\frac{1}{4}}\,,\quad m_s \sim g_s^{\frac{1}{2}} \, .
\end{eqaed}
These constraints are summarized in \cref{fig:bounds}. There is a window of parameter space in which all constraints are satisfied. However, a different implementation of the Higgs mechanism may be necessary: in the standard model, the scalar field at stake is in a fundamental representation, whereas scalar fields on brane worldvolumes are usually adjoint-valued. Nevertheless, the presence of unitary gauge groups and chiral fermions in this construction is encouraging to the effect of constructing more realistic models. In this respect, the type IIB construction of \cite{Danielsson:2022lsl} persists as a viable option \cite{Danielsson:2023alz}. It would be interesting to assess whether additional corrections produce a dynamical dark energy in that context, and what observational constraints would be.

\subsection{Early- and late-time cosmology} \label{sec:cosmo}

Having analyzed the viability of particle physics in our 0'B dark bubble model, one could ask is there is any viable realization of inflation. This would imply some sort of transition from an accelerated expansion to an era of radiation domination. This is highly non-trivial to understand from the perspective of this type of universes located on nucleating bubbles, since they lack a mechanism for inflation to end. In any case, let us focus on the 0'B dark bubble as a model of inflation, studying whether it can fit observational data. As showed in \cref{sec:0'B_dark bubblec}, the Hubble parameter $H$ contains a logarithmic dependence on the scale factor $a$. Thus it would be phenomenologically interesting to embed the inflationary paradigm in this scenario.

Consider a scalar field $\varphi$ with the quadratic potential we derived in \cref{sec:effective_scalars},
We now perform cosmological perturbation theory with the phantom scalar. To this end, we express the field as a background term, satisfying the Friedmann equations, plus a perturbation
\begin{eqaed}
    \phi(x,t)=\phi_0(t)+\dfrac{\delta\phi(t,x)}{a} \, .
\end{eqaed}
The linearized equation of motion for the perturbation is
\begin{eqaed}
    \delta\phi''+\left(k^2-\dfrac{a}{a''}\right)\delta\phi=0 \, ,
\end{eqaed}
where we have assumed a spatially flat gauge, as well as a slow-roll condition for the background field. We can immediately observe that this equation is independent on the sign of the kinetic energy. This allows us to straightforwardly recycle the usual results for scalar inflation, while keeping in mind the sign of the slow-roll parameter. At horizon crossing we find the corresponding scalar and tensor power spectra $P_\zeta, P_T$. For isotropic perturbations, the result is the same even if one chooses a different effective realization (e.g. vector field inflation), since the equations for the perturbations remain the same. The result is
\begin{eqaed}
    P_{\zeta}=\left(\dfrac{H^2}{2\pi \dot\phi_0^2}\right)^2=-\dfrac{H^2}{8\pi^2 \varepsilon} \, ,
\end{eqaed}
\begin{eqaed}
    P_{T}=2\dfrac{H^2}{\pi^2} \, .
\end{eqaed}
It is then straightforward to compute the spectral tilt
\begin{eqaed}
    n_s-1=\dfrac{d P_{\zeta}/dt}{H P_\zeta}=-2\epsilon-\eta
\end{eqaed}
and the tensor-to-scalar ratio
\begin{eqaed}
    r=\dfrac{P_T}{P_\zeta}=-16\epsilon \, .
\end{eqaed}
With the observational constrained values given by \cite{Planck:2018vyg}
\begin{eqaed}
    n_s-1=-0.035\pm0.004 \, , \qquad r<0.03 \, ,
\end{eqaed}
we can immediately infer that, unless $1\gg\eta\gg-\epsilon$, the scalar spectrum will be blue-shifted, corresponding to $n_s>1$, which is experimentally ruled out.  In order to relate the relevant quantities to observational data one must compute such quantities in terms of the number of e-folds $N_e$ (around 60) before the end of inflation. For the dark bubble, which leads to eternal inflation, one must introduce an additional mechanism for inflation to end. We keep the assumption of such mechanism general, and assume some field value $\phi_*$ corresponding to $N_e$ e-folds before the end of inflation. For the potential in \cref{eq:V_quadratic} we have 
\begin{eqaed}
    n_s-1=\frac{24 \left(3 \phi_* ^2-1\right)}{\left(3 \phi_* ^2+2\right)^2} \, , \qquad r=\frac{288 \phi_* ^2}{\left(3\phi_* ^2+2\right)^2} \, .
\end{eqaed}
In order to have red-shifted spectral tilt that approaches the experimental bounds one requires $\phi_*\approx 1/\sqrt{3}$, which in turn implies $r\gg1$. This also suggests that the slow-roll parameters are large, $|\epsilon|\gg0, |\eta|\gg 0$, which means that the expression considered is invalid, as the relations between the potential and the slow-roll parameters are derived within the slow-roll approximation.

Since the potential eq. \eqref{eq:V_quadratic} was computed without the assumption of slow-roll, the phantom scalar realization should be valid at all times, With this in mind, we can compute $\varepsilon$, $\eta$ (and with that $n_s$ and $r$) directly from the full expression for $a(\tau)$. We find
\begin{eqaed}
    \varepsilon= -\frac{\frac{\epsilon M^2}{2} }{(H_0+ \frac{\epsilon M^2}{2} \tau )^2} \, , \qquad \eta= -\frac{\epsilon M^2}{(H_0+ \frac{\epsilon M^2}{2} \tau )^2} \, .
\end{eqaed}
We can see that both slow-roll parameters are negative at all times, translating into a \emph{blue-shifted spectral tilt, incompatible with observations}.

We would also like to remark that the case of a LSBR scenario can be realized via a massive three-form field, given a potential with the appropriate behaviour. This sort of inflationary scenario can more generally arise from massive codimension-one forms, so it is perhaps not too surprising since we start from a massive four-form in $AdS_5$. For a massive three-form $C_3$, described by the action
\begin{eqaed}
     S = \int d^{4}x \sqrt{-g} \left(R - \frac{1}{g^2}F_{4}^2 -V(C_3^2) \right),
\end{eqaed}
the spectral tilt will be given by \cite{Koivisto:2009fb}
\begin{eqaed}
    n_s-1\simeq -2\varepsilon-\frac{3}{2}\eta \, .
\end{eqaed}
As such, the spectrum will also be blue-shifted. Results for both scalar and three-form realizations are shown in \cref{fig:cosmofigs} as a function of $\epsilon M^2/H_0$. Absolute scale invariance is recovered for $\epsilon=0$, as expected since one recovers pure \ac{dS} in this case.

\subsubsection*{Bounds on late time acceleration}

\begin{figure}
    \centering
    \begin{tabular}{c c}  % Two columns
        \includegraphics[width=0.43\textwidth]{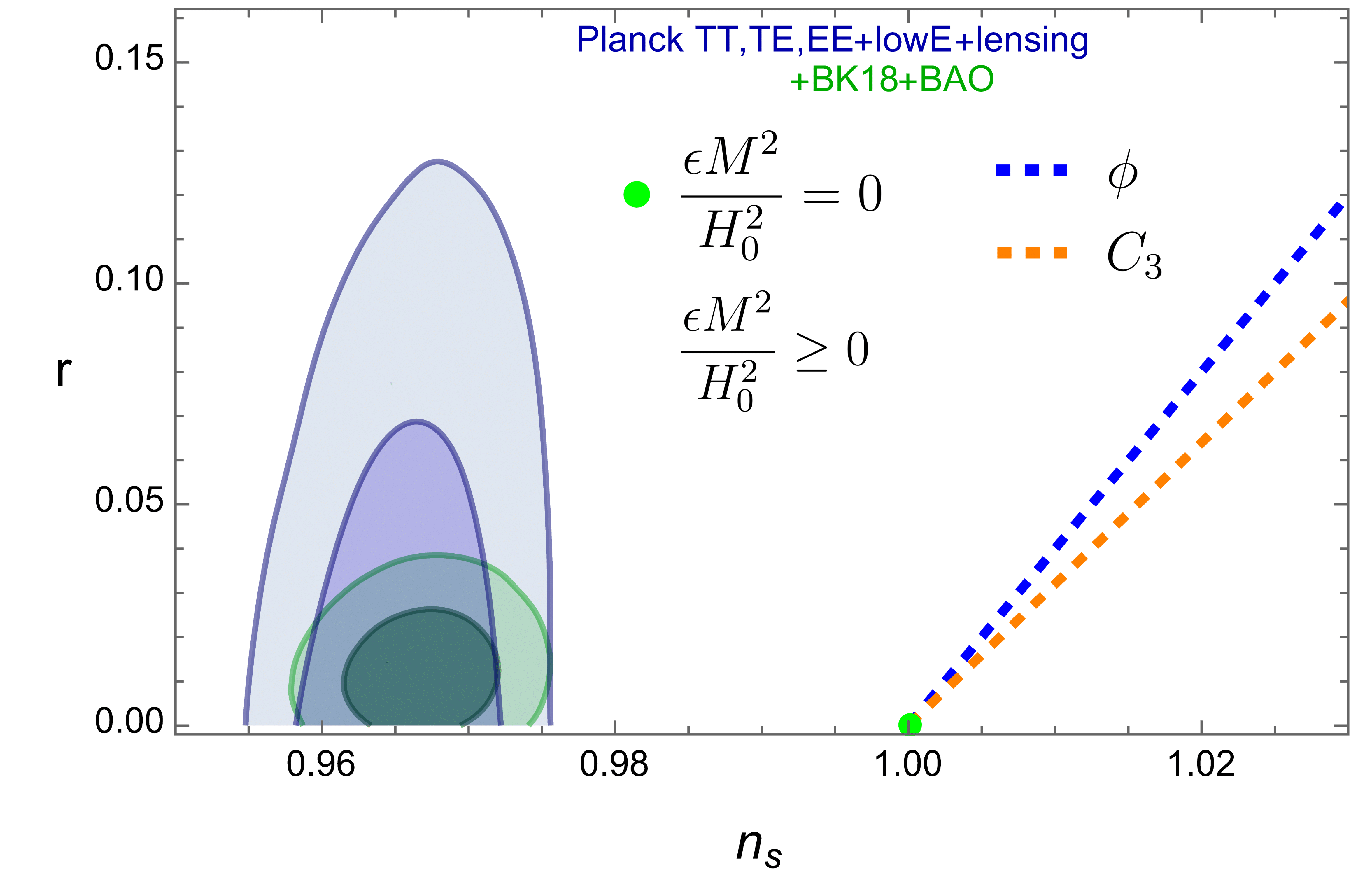} & 
        \includegraphics[width=0.45\textwidth]{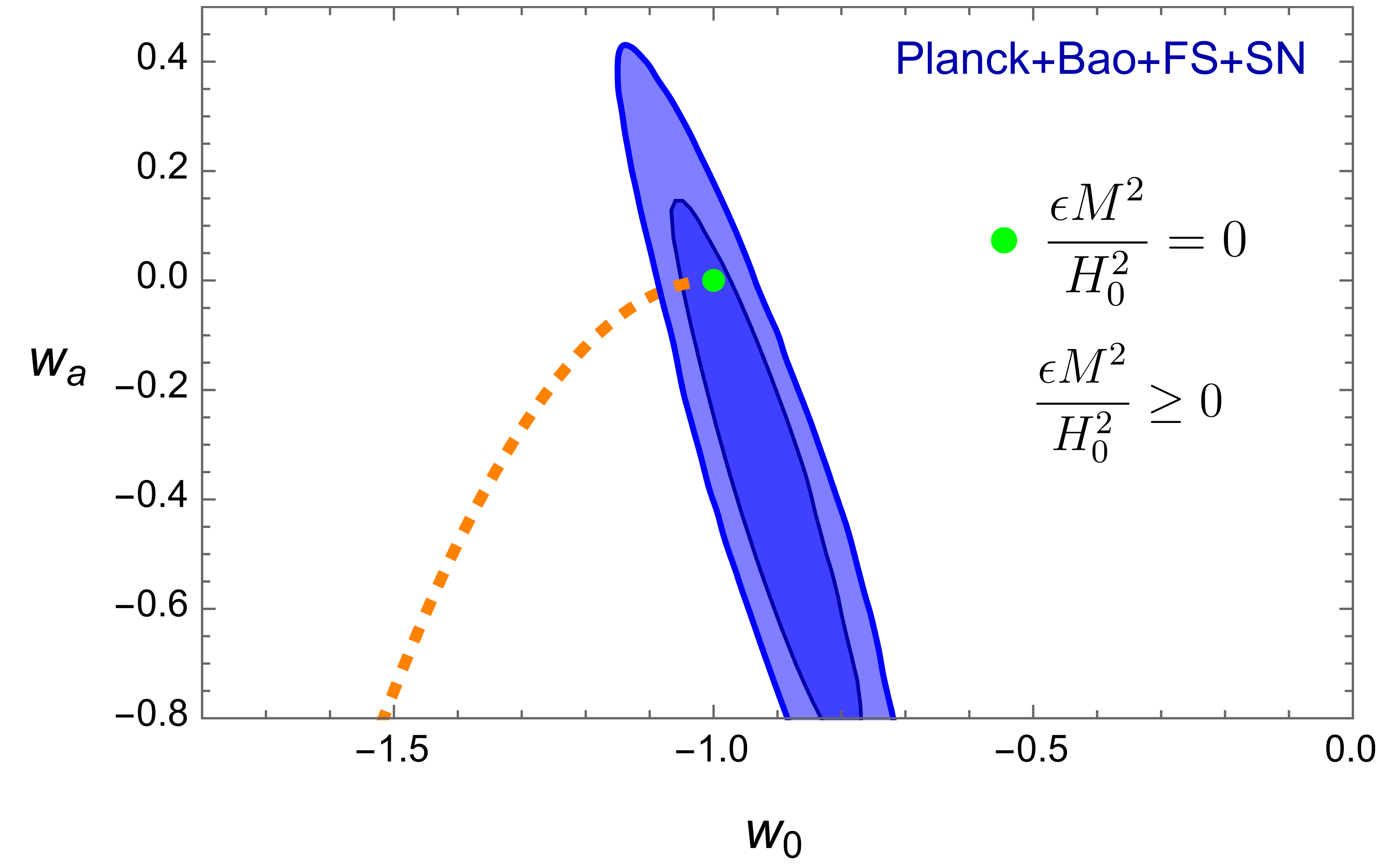} \\
    \end{tabular}
    \caption{ Left: Experimental bounds for inflation compared to the predictions of the model as a function of $\epsilon M^2/H_0^2$. Blue lines denote a phantom scalar realization, while the orange lines denote a realization via massive three-form fields. The points approaching $n_s\simeq1$ satisfy $\epsilon M^2\ll H_0^2$, but the spectrum is always blue-shifted. Right: Experimental bounds for late-time acceleration compared to observational constraints as a function of $\epsilon M^2/H_0^2$. For points that intersect the experimentally allowed region one has $H_0^2\gtrsim 0.1 \epsilon M^2$.}
    \label{fig:cosmofigs}
\end{figure}

We have already studied some bounds imposed by measurements of the cosmological constant of our universe in \cref{sec:bounds}. Here we impose bounds in terms of the equation of state for dark energy, which depend on the specific model. For time-independent dark energy ($w_0CDM$), the present-day equation of state parameter is constrained according to \cite{Planck:2018vyg} 
\begin{eqaed}
    w_0=-1.028\pm 0.031 \, .
\end{eqaed}
Perhaps a more interesting scenario is given by $w_0w_a CDM$, where the scale-dependence of dark energy can be parameterized as\footnote{It is worth mentioning a related parametrization at the level of \ac{EFT} scalar potentials \cite{VanRaamsdonk:2024sdp} which could be relevant to \ac{AdS} holography to cosmology.}
\begin{eqaed}
    w=w_0+(1-a)w_a
\end{eqaed}
with
\begin{eqaed}
    w_0=-0.957\pm0.08 \, ,\qquad w_a=-0.29^{+0.32}_{-0.26} \, .
\end{eqaed}
In the 0'B dark bubble model, appropriately choosing $H_0$ as the present day Hubble constant and $\tau=0$ as present time, we can expand the equation of state as
\begin{eqaed}
    w_{0'B}\simeq -1-\dfrac{\epsilon M^2}{3 H_0^2}-(1-a)\dfrac{\epsilon^2 M^4}{3 H_0^4} \, .
\end{eqaed}
In \cref{fig:cosmofigs} we plot this equation of state parameter against observational constraints. We find an agreement for $H_0^2 \gtrsim 0.1 \epsilon M^2$, within the region of validity of the \ac{EFT}.
Given the value of the Hubble parameter today,
\begin{eqaed}
H_0\approx 5.9\times10^{-61} M_\text{Pl},
\end{eqaed}
this imposes the bound
\begin{eqaed}
    \epsilon M^2 \lesssim 3.87\times 10^{-122} M_\text{Pl}^2 \, .
\end{eqaed}
In other words, $\epsilon M^2$ has to be, at most, of the order of the cosmological constant.
This can then be used to impose bounds on the string coupling $g_s$ and the number of stacked D3-branes sourcing the geometry, $N$, since
\begin{eqaed}
  \epsilon M^2 \approx g_s^2 M_\text{Pl}^2<10^{-122}M_\text{Pl}^2 \, .
\end{eqaed}
As a result, we obtain
\begin{eqaed}
    g_s\lesssim 10^{-61}\,,\quad\, N\gtrsim10^{61} \, .
\end{eqaed}
This sets the scales in the model as
\begin{eqaed}
    M \lesssim M_s \lesssim M_{10}\lesssim M_\text{Pl}\ll M_{5} \, ,
\end{eqaed}
with 
\begin{eqaed}
    M\lesssim \text{meV}\,,\quad\,M_{10}\lesssim \text{TeV}.
\end{eqaed}
Such scales could then in principle realize a braneworld analog of the dark dimension scenario, similarly to the results of \cite{Danielsson:2023alz}. The key difference in this work is the non-zero $\epsilon$, coming from the uncancelled dilaton tadpole of type 0'B string theory. This then strongly constraints the value of the string coupling, and makes it unfeasible to realize Standard Model gauge couplings on the brane. Differently from the results of \cite{Danielsson:2023alz}, this leads us to discard the 0'B dark bubble as a realistic model of our universe with the Standard Model located on the brane.

As an alternative toy model, we explore the possible scales coming from similar scenarios, but with the Standard Model localized on a stack of D$p$ branes with $p>3$. This requires replacing the $S^5$ with another suitable (Einstein) manifold with non-trivial homology, such that one can wrap the additional dimensions on a $(p-3)$-cycle $\Sigma_{p-3}$. In this case, the Lagrangian of \cref{eq:eff_dark bubbleI_action} is replaced by
\begin{eqaed}\label{eq:eff_dark bubbleI_action_Dp}
    \mathcal{L}_{\text{Dp}}\sim -T_\text{p} N_{p} \left( \frac{\alpha' \, U}{R(U)} \right)^4 e^{-\phi(U)} \tilde{R}(U)^{p-3}\sqrt{1 - \frac{\lambda \dot{U}^2}{U^4}}+ \mathcal{L}_{\text{CS}} \, .
\end{eqaed}
Since we still require the $\ads_5$ to be supported by five-form flux, we need to include the coupling between the wrapped D$p$ branes and $C_4$. Otherwise, the equations implied by energy conservation at large $U$ will be purely algebraic, and will not result in an inflationary cosmology. The coupling to the $C_4$ R-R potential is given by integrating over the compact directions in the Chern-Simons term
\begin{eqaed}
    S_\text{CS}\supset \mu_p N_p \int_{\mathbb{R}^{1,3}\times\Sigma_{p-3}} C_4\wedge e^{2\pi\alpha'\mathcal{F}} \, .
\end{eqaed}
For example, for $p=5,7$ respectively we have
\begin{eqaed}
  & \mu_5 N_5 (2\pi\alpha')\int_{\mathbb{R}^{1,3}\times\Sigma_{2}} C_4\wedge\mathcal{F} \, , \\
  & \frac{1}{2} \,\mu_7 N_7(2\pi\alpha')^2\int_{\mathbb{R}^{1,3}\times\Sigma_{4}} C_4\wedge\mathcal{F}\wedge\mathcal{F} \, ,
\end{eqaed}
where $\mathcal{F}$ is the field strength on the D$p$ brane stack. If this is not vanishing, a $C_4$ charge is induced on the brane according to
\begin{eqaed}
    Q_\text{eff}= \int_{\Sigma_{p-3}} e^{2\pi\alpha' \mathcal{F}} \, .
\end{eqaed}
The charge should then be chosen in a way such that the leading order contributions from the DBI and Chern-Simons actions cancel out, meaning
\begin{eqaed}
    Q_\text{eff}=\left(R_{\infty}^2\right)^{\frac{p-3}{2}} \, .
\end{eqaed}
In this case we can repeat the procedure detailed in \cref{sec:D3_0'B} to again obtain  LSBR cosmological solutions. From the Lagrangian we can directly infer the value of $\epsilon$ as
\begin{eqaed}
    \epsilon \propto g_s^2 N\, ,
\end{eqaed}
which is the same scaling as for D3-branes. In fact, by looking at the $p$-brane action, one could expect the dependence on $g_s, N$ to be generic since the leading contribution coming from \cref{eq:d3-branes_correction} is always of the form  $g_s^2 N$. Then, even ignoring all possible caveats of the model, building any construction of the 0'B dark bubble would lead to a phenomenology inconsistent with both inflation and late-time acceleration, as for the latter it would make it impossible to realize Standard Model gauge couplings on the braneworld.

\section{Conclusions} \label{sec:conclusions}

The dark bubble model is an intricate but dynamically natural construction which represents an alternative to the paradigm of standard compactification in string theory. It is a cosmological scenario which makes extensive use of elements from string theory, such as strings and branes, and can be embedded explicitly in type IIB and type 0'B strings. Whereas in the former case there is no conflict with phenomenology (at least thus far), the latter construction, albeit simple and elegant, is incompatible with cosmological observations and has to be discarded.

In this work, we found a general mechanism to embed electromagnetic gauge fields living on the brane even without the $B$-field, generalizing the results of \cite{Basile:2023tvh}. The two fields responsible for this interplay in general settings are the $B$-field and the R-R $C$-field. This represents a first step towards a complete description of the Standard Model in the dark bubble scenario. However, we have highlighted some potential obstruction to a full realization of the non-Abelian sector.

Moreover, we provided a no-go result for the effective description of D3-braneworlds arising from type 0'B strings. The evolution of the scale-factor $a(\tau)$ stems from a logarithmic dynamical dark energy, and indicates a quasi-\ac{dS} expansion. We showed that no effective descriptions with scalar fields at the two-derivative level can reproduce this running, although phantom scalars can. The prediction of a logarithmic dynamical dark energy seems promising for embedding inflation in the picture; however, we find a blue-shifted spectral tilt, incompatible with observations. Similarly, a realistic late-time cosmology would be realized together with extremely small couplings. On the other hand, a (semi)realistic particle physics may be consistently realized, including the correct gauge algebra, chiral fermions and a viable parameter space.

All in all, it remains important to assess which features of the model we are excluding are general and persist when seeking more elaborate constructions. At any rate, our results leave the type IIB realization of \cite{Danielsson:2022lsl} as the only existing proposal not excluded by observations, although more work is needed in order to accommodate the full Standard Model. This includes chiral fermions, the correct representations and Higgs mechanism, and non-Abelian gauge fields, whose minimal coupling to the induced gravity on the brane appears problematic. It would be interesting to investigate whether different couplings to gravity, as opposed to a standard \ac{EFT}, would produce experimental signatures to be tested in the near future.

Regardless of these considerations, the dark bubble scenario provides an example of a wholly complementary paradigm to compactification. Where breaking \ac{SUSY} and lifting moduli in compactifications entail several difficulties, the dark bubble requires the former and naturally implements the latter. The ubiquity of \ac{AdS} solutions and instabilities thereof upon breaking \ac{SUSY} turn from a catastrophe to a perk in this picture. Similarly, realizing scale separation and a positive dark energy become much more feasible relative to compactifications. As such, at the very least the dark bubble provides a novel perspective on the problems that string phenomenology ought to face, reminding us that novel angles to approach these difficult issues can be found in unexpected places.

\section*{Acknowledgements} \label{sec:acknowledgements}

I.B. is grateful to U. Danielsson, S. Giri, D. Panizo and V. Van Hemelryck for insightful discussions. J.M. acknowledges valuable exchanges with C. Kneißl and  A. Herraez. The work of I.B. is supported by the Origins Excellence Cluster and the German-Israel-Project (DIP) on Holography and the Swampland. A.B. thanks the Max Planck Institute for Physics for support and hospitality during this research project.

\printbibliography

@article{Bouhmadi-Lopez:2018lly,
    author = "Bouhmadi-L\'opez, Mariam and Brizuela, David and Garay, I\~naki",
    title = "{Quantum behavior of the ''Little Sibling'' of the Big Rip induced by a three-form field}",
    eprint = "1802.05164",
    archivePrefix = "arXiv",
    primaryClass = "gr-qc",
    doi = "10.1088/1475-7516/2018/09/031",
    journal = "JCAP",
    volume = "09",
    pages = "031",
    year = "2018"
}

@article{Kachru:2003aw,
	author = "Kachru, Shamit and Kallosh, Renata and Linde, Andrei D. and Trivedi, Sandip P.",
	title = "{De Sitter vacua in string theory}",
	eprint = "hep-th/0301240",
	archivePrefix = "arXiv",
	reportNumber = "SLAC-PUB-9630, SU-ITP-03-01, TIFR-TH-03-03",
	doi = "10.1103/PhysRevD.68.046005",
	journal = "Phys. Rev. D",
	volume = "68",
	pages = "046005",
	year = "2003"
}

@article{Giddings:2000mu,
	author = "Giddings, Steven B. and Katz, Emanuel and Randall, Lisa",
	title = "{Linearized gravity in brane backgrounds}",
	eprint = "hep-th/0002091",
	archivePrefix = "arXiv",
	reportNumber = "MIT-CTP-2944, PUPT-1913",
	doi = "10.1088/1126-6708/2000/03/023",
	journal = "JHEP",
	volume = "03",
	pages = "023",
	year = "2000"
}

@article{Karch:2000ct,
	author = "Karch, Andreas and Randall, Lisa",
	editor = "Duff, Michael J. and Liu, J.T. and Lu, J.",
	title = "{Locally localized gravity}",
	eprint = "hep-th/0011156",
	archivePrefix = "arXiv",
	reportNumber = "MIT-CTP-3099",
	doi = "10.1088/1126-6708/2001/05/008",
	journal = "JHEP",
	volume = "05",
	pages = "008",
	year = "2001"
}

@article{Banerjee:2018qey,
	author = "Banerjee, Souvik and Danielsson, Ulf and Dibitetto, Giuseppe and Giri, Suvendu and Schillo, Marjorie",
	title = "{Emergent de Sitter Cosmology from Decaying Anti--de Sitter Space}",
	eprint = "1807.01570",
	archivePrefix = "arXiv",
	primaryClass = "hep-th",
	reportNumber = "UUITP-27-18",
	doi = "10.1103/PhysRevLett.121.261301",
	journal = "Phys. Rev. Lett.",
	volume = "121",
	number = "26",
	pages = "261301",
	year = "2018"
}

@article{Banerjee:2020wix,
	author = "Banerjee, Souvik and Danielsson, Ulf and Giri, Suvendu",
	title = "{Dark bubbles: decorating the wall}",
	eprint = "2001.07433",
	archivePrefix = "arXiv",
	primaryClass = "hep-th",
	reportNumber = "UUITP-1/20",
	doi = "10.1007/JHEP04(2020)085",
	journal = "JHEP",
	volume = "20",
	pages = "085",
	year = "2020"
}

@article{Lust:2019zwm,
	author = "Lüst, Dieter and Palti, Eran and Vafa, Cumrun",
	title = "{AdS and the Swampland}",
	eprint = "1906.05225",
	archivePrefix = "arXiv",
	primaryClass = "hep-th",
	doi = "10.1016/j.physletb.2019.134867",
	journal = "Phys. Lett. B",
	volume = "797",
	pages = "134867",
	year = "2019"
}

@article{Basile:2020mpt,
    author = "Basile, Ivano and Lanza, Stefano",
    title = "{de Sitter in non-supersymmetric string theories: no-go theorems and brane-worlds}",
    eprint = "2007.13757",
    archivePrefix = "arXiv",
    primaryClass = "hep-th",
    doi = "10.1007/JHEP10(2020)108",
    journal = "JHEP",
    volume = "10",
    pages = "108",
    year = "2020"
}

@article{Antonelli:2019nar,
	author = "Antonelli, Riccardo and Basile, Ivano",
	title = "{Brane annihilation in non-supersymmetric strings}",
	eprint = "1908.04352",
	archivePrefix = "arXiv",
	primaryClass = "hep-th",
	doi = "10.1007/JHEP11(2019)021",
	journal = "JHEP",
	volume = "11",
	pages = "021",
	year = "2019"
}

@article{Basile:2018irz,
      author         = "Basile, I. and Mourad, J. and Sagnotti, A.",
      title          = "{On Classical Stability with Broken Supersymmetry}",
      journal        = "JHEP",
      volume         = "01",
      year           = "2019",
      pages          = "174",
      doi            = "10.1007/JHEP01(2019)174",
      eprint         = "1811.11448",
      archivePrefix  = "arXiv",
      primaryClass   = "hep-th",
      SLACcitation   = "%%CITATION = ARXIV:1811.11448;%%"
}

@article{Maldacena:1997re,
      author         = "Maldacena, Juan Martin",
      title          = "{The Large N limit of superconformal field theories and
                        supergravity}",
      journal        = "Int. J. Theor. Phys.",
      volume         = "38",
      year           = "1999",
      pages          = "1113-1133",
      doi            = "10.1023/A:1026654312961, 10.4310/ATMP.1998.v2.n2.a1",
      note           = "[Adv. Theor. Math. Phys.2,231(1998)]",
      eprint         = "hep-th/9711200",
      archivePrefix  = "arXiv",
      primaryClass   = "hep-th",
      reportNumber   = "HUTP-97-A097, HUTP-98-A097",
      SLACcitation   = "%%CITATION = HEP-TH/9711200;%%"
}

@article{AlvarezGaume:1986jb,
      author         = "Alvarez-Gaume, Luis and Ginsparg, Paul H. and Moore,
                        Gregory W. and Vafa, C.",
      title          = "{An O(16) x O(16) Heterotic String}",
      journal        = "Phys. Lett.",
      volume         = "B171",
      year           = "1986",
      pages          = "155-162",
      doi            = "10.1016/0370-2693(86)91524-8",
      reportNumber   = "HUTP-86/A013",
      SLACcitation   = "%%CITATION = PHLTA,B171,155;%%"
}

@article{Dixon:1986iz,
      author         = "Dixon, Lance J. and Harvey, Jeffrey A.",
      title          = "{String Theories in Ten-Dimensions Without Space-Time
                        Supersymmetry}",
      journal        = "Nucl. Phys.",
      volume         = "B274",
      year           = "1986",
      pages          = "93-105",
      doi            = "10.1016/0550-3213(86)90619-X",
      note           = "[,93(1986)]",
      reportNumber   = "PRINT-86-0244 (PRINCETON)",
      SLACcitation   = "%%CITATION = NUPHA,B274,93;%%"
}

@inproceedings{Sagnotti:1995ga,
      author         = "Sagnotti, Augusto",
      title          = "{Some properties of open string theories}",
      booktitle      = "{Supersymmetry and unification of fundamental
                        interactions. Proceedings, International Workshop, SUSY
                        95, Palaiseau, France, May 15-19}",
      year           = "1995",
      pages          = "473-484",
      eprint         = "hep-th/9509080",
      archivePrefix  = "arXiv",
      primaryClass   = "hep-th",
      reportNumber   = "ROM2F-95-18",
      SLACcitation   = "%%CITATION = HEP-TH/9509080;%%"
}

@article{Sagnotti:1996qj,
      author         = "Sagnotti, Augusto",
      title          = "{Surprises in open string perturbation theory}",
      booktitle      = "{Theory of elementary particles. Proceedings, 30th
                        International Symposium Ahrenshoop, Buckow, Germany,
                        August 27-31, 1996}",
      journal        = "Nucl. Phys. Proc. Suppl.",
      volume         = "56B",
      year           = "1997",
      pages          = "332-343",
      doi            = "10.1016/S0920-5632(97)00344-7",
      eprint         = "hep-th/9702093",
      archivePrefix  = "arXiv",
      primaryClass   = "hep-th",
      reportNumber   = "ROM2F-97-4",
      SLACcitation   = "%%CITATION = HEP-TH/9702093;%%"
}

@article{Sugimoto:1999tx,
      author         = "Sugimoto, Shigeki",
      title          = "{Anomaly cancellations in type I D-9 - anti-D-9 system
                        and the USp(32) string theory}",
      journal        = "Prog. Theor. Phys.",
      volume         = "102",
      year           = "1999",
      pages          = "685-699",
      doi            = "10.1143/PTP.102.685",
      eprint         = "hep-th/9905159",
      archivePrefix  = "arXiv",
      primaryClass   = "hep-th",
      reportNumber   = "YITP-99-25",
      SLACcitation   = "%%CITATION = HEP-TH/9905159;%%"
}

@article{Antoniadis:1999xk,
      author         = "Antoniadis, Ignatios and Dudas, E. and Sagnotti, A.",
      title          = "{Brane supersymmetry breaking}",
      journal        = "Phys. Lett.",
      volume         = "B464",
      year           = "1999",
      pages          = "38-45",
      doi            = "10.1016/S0370-2693(99)01023-0",
      eprint         = "hep-th/9908023",
      archivePrefix  = "arXiv",
      primaryClass   = "hep-th",
      reportNumber   = "CPHT-S727-0799, LPT-ORSAY-99-60, ROM2F-99-23",
      SLACcitation   = "%%CITATION = HEP-TH/9908023;%%"
}

@article{Angelantonj:1999jh,
      author         = "Angelantonj, Carlo",
      title          = "{Comments on open string orbifolds with a nonvanishing
                        B(ab)}",
      journal        = "Nucl. Phys.",
      volume         = "B566",
      year           = "2000",
      pages          = "126-150",
      doi            = "10.1016/S0550-3213(99)00662-8",
      eprint         = "hep-th/9908064",
      archivePrefix  = "arXiv",
      primaryClass   = "hep-th",
      reportNumber   = "CPHT-S718-0599, LPTENS-99-27, CPHT-718.0599",
      SLACcitation   = "%%CITATION = HEP-TH/9908064;%%"
}

@article{Aldazabal:1999jr,
      author         = "Aldazabal, G. and Uranga, A. M.",
      title          = "{Tachyon free nonsupersymmetric type IIB orientifolds via
                        Brane - anti-brane systems}",
      journal        = "JHEP",
      volume         = "10",
      year           = "1999",
      pages          = "024",
      doi            = "10.1088/1126-6708/1999/10/024",
      eprint         = "hep-th/9908072",
      archivePrefix  = "arXiv",
      primaryClass   = "hep-th",
      reportNumber   = "CAB-IB-2911299, IASSNS-HEP-99-79",
      SLACcitation   = "%%CITATION = HEP-TH/9908072;%%"
}

@article{Angelantonj:1999ms,
      author         = "Angelantonj, C. and Antoniadis, Ignatios and
                        D'Appollonio, G. and Dudas, E. and Sagnotti, A.",
      title          = "{Type I vacua with brane supersymmetry breaking}",
      journal        = "Nucl. Phys.",
      volume         = "B572",
      year           = "2000",
      pages          = "36-70",
      doi            = "10.1016/S0550-3213(00)00052-3",
      eprint         = "hep-th/9911081",
      archivePrefix  = "arXiv",
      primaryClass   = "hep-th",
      reportNumber   = "LPTENS-99-38, CPTH-S743-1099, NSF-ITP-99-127,
                        DFF-347-10-99, LPT-ORSAY-99-80, ROM2F-99-40,
                        CPTH-S743.1099, --LPT-ORSAY-99-80",
      SLACcitation   = "%%CITATION = HEP-TH/9911081;%%"
}

@article{Dudas:2000nv,
      author         = "Dudas, E. and Mourad, J.",
      title          = "{Consistent gravitino couplings in nonsupersymmetric
                        strings}",
      journal        = "Phys. Lett.",
      volume         = "B514",
      year           = "2001",
      pages          = "173-182",
      doi            = "10.1016/S0370-2693(01)00777-8",
      eprint         = "hep-th/0012071",
      archivePrefix  = "arXiv",
      primaryClass   = "hep-th",
      reportNumber   = "LPT-ORSAY-00-128",
      SLACcitation   = "%%CITATION = HEP-TH/0012071;%%"
}

@article{Dudas:2000ff,
      author         = "Dudas, E. and Mourad, J.",
      title          = "{Brane solutions in strings with broken supersymmetry and
                        dilaton tadpoles}",
      journal        = "Phys. Lett.",
      volume         = "B486",
      year           = "2000",
      pages          = "172-178",
      doi            = "10.1016/S0370-2693(00)00734-6",
      eprint         = "hep-th/0004165",
      archivePrefix  = "arXiv",
      primaryClass   = "hep-th",
      reportNumber   = "LPT-ORSAY-00-43, LPTM-00-25, LPT-00-43",
      SLACcitation   = "%%CITATION = HEP-TH/0004165;%%"
}

@article{Dudas:2001wd,
      author         = "Dudas, E. and Mourad, J. and Sagnotti, A.",
      title          = "{Charged and uncharged D-branes in various string
                        theories}",
      journal        = "Nucl. Phys.",
      volume         = "B620",
      year           = "2002",
      pages          = "109-151",
      doi            = "10.1016/S0550-3213(01)00552-1",
      eprint         = "hep-th/0107081",
      archivePrefix  = "arXiv",
      primaryClass   = "hep-th",
      reportNumber   = "LPT-ORSAY-01-56, ROM2F-01-18",
      SLACcitation   = "%%CITATION = HEP-TH/0107081;%%"
}

@article{Angelantonj:2002ct,
      author         = "Angelantonj, Carlo and Sagnotti, Augusto",
      title          = "{Open strings}",
      journal        = "Phys. Rept.",
      volume         = "371",
      year           = "2002",
      pages          = "1-150",
      doi            = "10.1016/S0370-1573(02)00273-9,
                        10.1016/S0370-1573(03)00006-1",
      note           = "[Erratum: Phys. Rept.376,no.6,407(2003)]",
      eprint         = "hep-th/0204089",
      archivePrefix  = "arXiv",
      primaryClass   = "hep-th",
      reportNumber   = "CERN-TH-2002-025, ROM2F-2002-08, LPTENS-02-14,
                        CPHT-RR-020-0202, CPHT-RR-020.0202",
      SLACcitation   = "%%CITATION = HEP-TH/0204089;%%"
}

@article{Mourad:2016xbk,
      author         = "Mourad, J. and Sagnotti, A.",
      title          = "{$AdS$ Vacua from Dilaton Tadpoles and Form Fluxes}",
      journal        = "Phys. Lett.",
      volume         = "B768",
      year           = "2017",
      pages          = "92-96",
      doi            = "10.1016/j.physletb.2017.02.053",
      eprint         = "1612.08566",
      archivePrefix  = "arXiv",
      primaryClass   = "hep-th",
      SLACcitation   = "%%CITATION = ARXIV:1612.08566;%%"
}

@article{Angelantonj:2000kh,
      author         = "Angelantonj, Carlo and Armoni, Adi",
      title          = "{RG flow, Wilson loops and the dilaton tadpole}",
      journal        = "Phys. Lett.",
      volume         = "B482",
      year           = "2000",
      pages          = "329-336",
      doi            = "10.1016/S0370-2693(00)00475-5",
      eprint         = "hep-th/0003050",
      archivePrefix  = "arXiv",
      primaryClass   = "hep-th",
      reportNumber   = "CPHT-S004-0300, LPTENS-00-07",
      SLACcitation   = "%%CITATION = HEP-TH/0003050;%%"
}

@article{Angelantonj:1999qg,
      author         = "Angelantonj, Carlo and Armoni, Adi",
      title          = "{Nontachyonic type 0B orientifolds, nonsupersymmetric
                        gauge theories and cosmological RG flow}",
      journal        = "Nucl. Phys.",
      volume         = "B578",
      year           = "2000",
      pages          = "239-258",
      doi            = "10.1016/S0550-3213(00)00136-X",
      eprint         = "hep-th/9912257",
      archivePrefix  = "arXiv",
      primaryClass   = "hep-th",
      reportNumber   = "CPTH-S757-1299, LPTENS-99-55",
      SLACcitation   = "%%CITATION = HEP-TH/9912257;%%"
}

@article{Dudas:2000sn,
      author         = "Dudas, E. and Mourad, J.",
      title          = "{D-branes in nontachyonic 0B orientifolds}",
      journal        = "Nucl. Phys.",
      volume         = "B598",
      year           = "2001",
      pages          = "189-224",
      doi            = "10.1016/S0550-3213(00)00781-1",
      eprint         = "hep-th/0010179",
      archivePrefix  = "arXiv",
      primaryClass   = "hep-th",
      reportNumber   = "LPT-ORSAY-00-81",
      SLACcitation   = "%%CITATION = HEP-TH/0010179;%%"
}

@article{Klebanov:1998yya,
      author         = "Klebanov, Igor R. and Tseytlin, Arkady A.",
      title          = "{D-branes and dual gauge theories in type 0 strings}",
      journal        = "Nucl. Phys.",
      volume         = "B546",
      year           = "1999",
      pages          = "155-181",
      doi            = "10.1016/S0550-3213(99)00041-3",
      eprint         = "hep-th/9811035",
      archivePrefix  = "arXiv",
      primaryClass   = "hep-th",
      reportNumber   = "PUPT-1819, IMPERIAL-TP-98-99-07",
      SLACcitation   = "%%CITATION = HEP-TH/9811035;%%"
}

@article{ArkaniHamed:2006dz,
      author         = "Arkani-Hamed, Nima and Motl, Lubos and Nicolis, Alberto
                        and Vafa, Cumrun",
      title          = "{The String landscape, black holes and gravity as the
                        weakest force}",
      journal        = "JHEP",
      volume         = "06",
      year           = "2007",
      pages          = "060",
      doi            = "10.1088/1126-6708/2007/06/060",
      eprint         = "hep-th/0601001",
      archivePrefix  = "arXiv",
      primaryClass   = "hep-th",
      reportNumber   = "HUTP-05-A0057",
      SLACcitation   = "%%CITATION = HEP-TH/0601001;%%"
}

@article{Ooguri:2016pdq,
      author         = "Ooguri, Hirosi and Vafa, Cumrun",
      title          = "{Non-supersymmetric AdS and the Swampland}",
      year           = "2016",
      eprint         = "1610.01533",
      archivePrefix  = "arXiv",
      primaryClass   = "hep-th",
      reportNumber   = "CALT-TH-2016-027, IPMU16-0139",
      SLACcitation   = "%%CITATION = ARXIV:1610.01533;%%"
}

@article{Palti:2019pca,
      author         = "Palti, Eran",
      title          = "{The Swampland: Introduction and Review}",
      journal        = "Fortsch. Phys.",
      volume         = "67",
      year           = "2019",
      number         = "6",
      pages          = "1900037",
      doi            = "10.1002/prop.201900037",
      eprint         = "1903.06239",
      archivePrefix  = "arXiv",
      primaryClass   = "hep-th",
      reportNumber   = "MPP-2019-53",
      SLACcitation   = "%%CITATION = ARXIV:1903.06239;%%"
}

@article{Bedroya:2019snp,
	author = "Bedroya, Alek and Vafa, Cumrun",
	title = "{Trans-Planckian Censorship and the Swampland}",
	eprint = "1909.11063",
	archivePrefix = "arXiv",
	primaryClass = "hep-th",
	month = "9",
	year = "2019"
}

@article{Ooguri:2006in,
    author = "Ooguri, Hirosi and Vafa, Cumrun",
    title = "{On the Geometry of the String Landscape and the Swampland}",
    eprint = "hep-th/0605264",
    archivePrefix = "arXiv",
    reportNumber = "CALT-68-2600, HUTP-06-A017",
    doi = "10.1016/j.nuclphysb.2006.10.033",
    journal = "Nucl. Phys. B",
    volume = "766",
    pages = "21--33",
    year = "2007"
}

@article{Montero:2022prj,
    author = "Montero, Miguel and Vafa, Cumrun and Valenzuela, Irene",
    title = "{The Dark Dimension and the Swampland}",
    eprint = "2205.12293",
    archivePrefix = "arXiv",
    primaryClass = "hep-th",
    month = "5",
    year = "2022"
}

@article{Danielsson:2021tyb,
    author = "Danielsson, U. H. and Panizo, D. and Tielemans, R. and Van Riet, T.",
    title = "{Higher-dimensional view on quantum cosmology}",
    eprint = "2105.03253",
    archivePrefix = "arXiv",
    primaryClass = "hep-th",
    reportNumber = "UUITP - 22/21",
    doi = "10.1103/PhysRevD.104.086015",
    journal = "Phys. Rev. D",
    volume = "104",
    number = "8",
    pages = "086015",
    year = "2021"
}

@article{Banerjee:2021yrb,
    author = "Banerjee, Souvik and Danielsson, Ulf and Giri, Suvendu",
    title = "{Curing with Hemlock: Escaping the swampland using instabilities from string theory}",
    eprint = "2103.17121",
    archivePrefix = "arXiv",
    primaryClass = "hep-th",
    reportNumber = "UUITP-15/21",
    doi = "10.1142/S0218271821420293",
    journal = "Int. J. Mod. Phys. D",
    volume = "30",
    number = "14",
    pages = "2142029",
    year = "2021"
}

@article{Banerjee:2019fzz,
    author = "Banerjee, Souvik and Danielsson, Ulf and Dibitetto, Giuseppe and Giri, Suvendu and Schillo, Marjorie",
    title = "{de Sitter Cosmology on an expanding bubble}",
    eprint = "1907.04268",
    archivePrefix = "arXiv",
    primaryClass = "hep-th",
    reportNumber = "UUITP-26/19",
    doi = "10.1007/JHEP10(2019)164",
    journal = "JHEP",
    volume = "10",
    pages = "164",
    year = "2019"
}

@article{van_Beest_2022,
   title={Lectures on the Swampland Program in String Compactifications},
   volume={989},
   ISSN={0370-1573},
   url={http://dx.doi.org/10.1016/j.physrep.2022.09.002},
   DOI={10.1016/j.physrep.2022.09.002},
   journal={Physics Reports},
   publisher={Elsevier BV},
   author={van Beest, Marieke and Calderón-Infante, José and Mirfendereski, Delaram and Valenzuela, Irene},
   year={2022},
   month=nov, pages={1–50} }

@misc{agmon2023lecturesstringlandscapeswampland,
      title={Lectures on the string landscape and the Swampland}, 
      author={Nathan Benjamin Agmon and Alek Bedroya and Monica Jinwoo Kang and Cumrun Vafa},
      year={2023},
      eprint={2212.06187},
      archivePrefix={arXiv},
      primaryClass={hep-th},
      url={https://arxiv.org/abs/2212.06187}, 
}

@article{Randall_1999,
   title={An Alternative to Compactification},
   volume={83},
   ISSN={1079-7114},
   url={http://dx.doi.org/10.1103/PhysRevLett.83.4690},
   DOI={10.1103/physrevlett.83.4690},
   number={23},
   journal={Physical Review Letters},
   publisher={American Physical Society (APS)},
   author={Randall, Lisa and Sundrum, Raman},
   year={1999},
   month=dec, pages={4690–4693} }

@article{Banerjee_2019,
   title={De Sitter cosmology on an expanding bubble},
   volume={2019},
   ISSN={1029-8479},
   url={http://dx.doi.org/10.1007/JHEP10(2019)164},
   DOI={10.1007/jhep10(2019)164},
   number={10},
   journal={Journal of High Energy Physics},
   publisher={Springer Science and Business Media LLC},
   author={Banerjee, Souvik and Danielsson, Ulf and Dibitetto, Giuseppe and Giri, Suvendu and Schillo, Marjorie},
   year={2019},
   month=oct }

@misc{danielsson2024chargednariaiblackholes,
      title={Charged Nariai black holes on the dark bubble}, 
      author={Ulf Danielsson and Vincent Van Hemelryck},
      year={2024},
      eprint={2405.13679},
      archivePrefix={arXiv},
      primaryClass={hep-th},
      url={https://arxiv.org/abs/2405.13679}, 
}

@article{Banerjee:2022ree,
    author = "Banerjee, Souvik and Danielsson, Ulf and Giri, Suvendu",
    title = "{Features of a dark energy model in string theory}",
    eprint = "2212.14004",
    archivePrefix = "arXiv",
    primaryClass = "hep-th",
    reportNumber = "UUITP-63/22",
    doi = "10.1103/PhysRevD.108.126009",
    journal = "Phys. Rev. D",
    volume = "108",
    number = "12",
    pages = "126009",
    year = "2023"
}

@article{Danielsson:2022fhd,
    author = "Danielsson, Ulf and Panizo, Daniel and Tielemans, Rob",
    title = "{Gravitational waves in dark bubble cosmology}",
    eprint = "2202.00545",
    archivePrefix = "arXiv",
    primaryClass = "hep-th",
    reportNumber = "UUITP - 03/22",
    doi = "10.1103/PhysRevD.106.024002",
    journal = "Phys. Rev. D",
    volume = "106",
    number = "2",
    pages = "024002",
    year = "2022"
}

@misc{dvali2010speciesstrings,
      title={Species and Strings}, 
      author={Gia Dvali and Cesar Gomez},
      year={2010},
      eprint={1004.3744},
      archivePrefix={arXiv},
      primaryClass={hep-th},
      url={https://arxiv.org/abs/1004.3744}, 
}

@article{Banerjee:2023uto,
    author = "Banerjee, Souvik and Danielsson, Ulf and Zemsch, Maximilian",
    title = "{The dark bubbleography}",
    eprint = "2311.16242",
    archivePrefix = "arXiv",
    primaryClass = "hep-th",
    doi = "10.1007/JHEP02(2024)102",
    journal = "JHEP",
    volume = "02",
    pages = "102",
    year = "2024"
}

@article{Basile_2021,
   title={Supersymmetry breaking, brane dynamics and Swampland conjectures},
   volume={2021},
   ISSN={1029-8479},
   url={http://dx.doi.org/10.1007/JHEP10(2021)080},
   DOI={10.1007/jhep10(2021)080},
   number={10},
   journal={Journal of High Energy Physics},
   publisher={Springer Science and Business Media LLC},
   author={Basile, Ivano},
   year={2021},
   month=oct }

@article{Israel:1966rt,
    author = "Israel, W.",
    title = "{Singular hypersurfaces and thin shells in general relativity}",
    doi = "10.1007/BF02710419",
    journal = "Nuovo Cim. B",
    volume = "44S10",
    pages = "1",
    year = "1966",
    note = "[Erratum: Nuovo Cim.B 48, 463 (1967)]"
}

@article{Planck:2018vyg,
    author = "Aghanim, N. and others",
    collaboration = "Planck",
    title = "{Planck 2018 results. VI. Cosmological parameters}",
    eprint = "1807.06209",
    archivePrefix = "arXiv",
    primaryClass = "astro-ph.CO",
    doi = "10.1051/0004-6361/201833910",
    journal = "Astron. Astrophys.",
    volume = "641",
    pages = "A6",
    year = "2020",
    note = "[Erratum: Astron.Astrophys. 652, C4 (2021)]"
}

@article{Koivisto:2009fb,
    author = "Koivisto, Tomi S. and Nunes, Nelson J.",
    title = "{Inflation and dark energy from three-forms}",
    eprint = "0908.0920",
    archivePrefix = "arXiv",
    primaryClass = "astro-ph.CO",
    doi = "10.1103/PhysRevD.80.103509",
    journal = "Phys. Rev. D",
    volume = "80",
    pages = "103509",
    year = "2009"
}

@article{Danielsson:2023alz,
    author = "Danielsson, Ulf and Panizo, Daniel",
    title = "{Experimental tests of dark bubble cosmology}",
    eprint = "2311.14589",
    archivePrefix = "arXiv",
    primaryClass = "hep-th",
    doi = "10.1103/PhysRevD.109.026003",
    journal = "Phys. Rev. D",
    volume = "109",
    number = "2",
    pages = "026003",
    year = "2024"
}

@article{Danielsson:2022lsl,
    author = "Danielsson, Ulf and Henriksson, Oscar and Panizo, Daniel",
    title = "{Stringy realization of a small and positive cosmological constant in dark bubble cosmology}",
    eprint = "2211.10191",
    archivePrefix = "arXiv",
    primaryClass = "hep-th",
    reportNumber = "UUITP - 49/22, HIP-2022-29/TH",
    doi = "10.1103/PhysRevD.107.026020",
    journal = "Phys. Rev. D",
    volume = "107",
    number = "2",
    pages = "026020",
    year = "2023"
}

@article{Albarran:2016mdu,
    author = "Albarran, Imanol and Bouhmadi-L\'opez, Mariam and Morais, Jo\~ao",
    title = "{Cosmological perturbations in an effective and genuinely phantom dark energy Universe}",
    eprint = "1611.00392",
    archivePrefix = "arXiv",
    primaryClass = "astro-ph.CO",
    doi = "10.1016/j.dark.2017.04.002",
    journal = "Phys. Dark Univ.",
    volume = "16",
    pages = "94--108",
    year = "2017"
}

@article{Basile:2023tvh,
    author = "Basile, Ivano and Danielsson, Ulf and Giri, Suvendu and Panizo, Daniel",
    title = "{Shedding light on dark bubble cosmology}",
    eprint = "2310.15032",
    archivePrefix = "arXiv",
    primaryClass = "hep-th",
    reportNumber = "UUITP-30/23",
    doi = "10.1007/JHEP02(2024)112",
    journal = "JHEP",
    volume = "02",
    pages = "112",
    year = "2024"
}

@misc{obied2018sitterspaceswampland,
      title={De Sitter Space and the Swampland}, 
      author={Georges Obied and Hirosi Ooguri and Lev Spodyneiko and Cumrun Vafa},
      year={2018},
      eprint={1806.08362},
      archivePrefix={arXiv},
      primaryClass={hep-th},
      url={https://arxiv.org/abs/1806.08362}, 
}

@article{Anastasi:2025puv,
    author = "Anastasi, Edoardo and Angius, Roberta and Huertas, Jes\'us and Uranga, Angel and Wang, Chuying",
    title = "{Relative Quantum Gravity: Localized Gravity and the Swampland}",
    eprint = "2501.03310",
    archivePrefix = "arXiv",
    primaryClass = "hep-th",
    month = "1",
    year = "2025"
}

@article{Schellekens:1986xh,
    author = "Schellekens, A. N. and Warner, N. P.",
    title = "{Anomalies, Characters and Strings}",
    reportNumber = "CERN-TH-4529/86",
    doi = "10.1016/0550-3213(87)90108-8",
    journal = "Nucl. Phys. B",
    volume = "287",
    pages = "317",
    year = "1987"
}

@article{Hanany:2010da,
    author = "Hanany, Amihay and Forcella, Davide and Troost, Jan",
    title = "{The Covariant perturbative string spectrum}",
    eprint = "1007.2622",
    archivePrefix = "arXiv",
    primaryClass = "hep-th",
    doi = "10.1016/j.nuclphysb.2011.01.002",
    journal = "Nucl. Phys. B",
    volume = "846",
    pages = "212--225",
    year = "2011"
}

@article{Larotonda:2024thv,
    author = "Larotonda, Vittorio and Lin, Ling",
    title = "{Anomaly Inflow and Gauge Group Topology in the 10d Sugimoto String Theory}",
    eprint = "2412.17894",
    archivePrefix = "arXiv",
    primaryClass = "hep-th",
    month = "12",
    year = "2024"
}

@article{BoyleSmith:2023xkd,
    author = "Boyle Smith, Philip and Lin, Ying-Hsuan and Tachikawa, Yuji and Zheng, Yunqin",
    title = "{Classification of chiral fermionic CFTs of central charge $\le$ 16}",
    eprint = "2303.16917",
    archivePrefix = "arXiv",
    primaryClass = "hep-th",
    doi = "10.21468/SciPostPhys.16.2.058",
    journal = "SciPost Phys.",
    volume = "16",
    number = "2",
    pages = "058",
    year = "2024"
}

@article{McInnes:1999va,
    author = "McInnes, Brett",
    title = "{The Semispin groups in string theory}",
    eprint = "hep-th/9906059",
    archivePrefix = "arXiv",
    doi = "10.1063/1.532999",
    journal = "J. Math. Phys.",
    volume = "40",
    pages = "4699--4712",
    year = "1999"
}

@article{Basile:2023knk,
    author = "Basile, Ivano and Debray, Arun and Delgado, Matilda and Montero, Miguel",
    title = "{Global anomalies \& bordism of non-supersymmetric strings}",
    eprint = "2310.06895",
    archivePrefix = "arXiv",
    primaryClass = "hep-th",
    reportNumber = "IFT-23-129",
    doi = "10.1007/JHEP02(2024)092",
    journal = "JHEP",
    volume = "02",
    pages = "092",
    year = "2024"
}

@article{Kneissl:2024zox,
    author = "Kneissl, Christian",
    title = "{Spin cobordism and the gauge group of type I/heterotic string theory}",
    eprint = "2407.20333",
    archivePrefix = "arXiv",
    primaryClass = "hep-th",
    reportNumber = "MPP-2024-159",
    doi = "10.1007/JHEP01(2025)181",
    journal = "JHEP",
    volume = "01",
    pages = "181",
    year = "2025"
}

@article{Nakajima:2023zsh,
    author = "Nakajima, Sota",
    title = "{New non-supersymmetric heterotic string theory with reduced rank and exponential suppression of the cosmological constant}",
    eprint = "2303.04489",
    archivePrefix = "arXiv",
    primaryClass = "hep-th",
    reportNumber = "KEK-TH-2503",
    month = "3",
    year = "2023"
}

@article{Angelantonj:2024jtu,
    author = "Angelantonj, Carlo and Florakis, Ioannis and Leone, Giorgio and Perugini, Diego",
    title = "{Non-supersymmetric non-tachyonic heterotic vacua with reduced rank in various dimensions}",
    eprint = "2407.09597",
    archivePrefix = "arXiv",
    primaryClass = "hep-th",
    doi = "10.1007/JHEP10(2024)216",
    journal = "JHEP",
    volume = "10",
    pages = "216",
    year = "2024"
}

@article{Mourad:2021roa,
    author = "Mourad, J. and Sagnotti, A.",
    title = "{On warped string vacuum profiles and cosmologies. Part II. Non-supersymmetric strings}",
    eprint = "2109.12328",
    archivePrefix = "arXiv",
    primaryClass = "hep-th",
    doi = "10.1007/JHEP12(2021)138",
    journal = "JHEP",
    volume = "12",
    pages = "138",
    year = "2021"
}

@article{Mourad:2024dur,
    author = "Mourad, J. and Raucci, S. and Sagnotti, A.",
    title = "{Brane-like solutions and other non-supersymmetric vacua}",
    eprint = "2406.14926",
    archivePrefix = "arXiv",
    primaryClass = "hep-th",
    doi = "10.1007/JHEP10(2024)054",
    journal = "JHEP",
    volume = "10",
    pages = "054",
    year = "2024"
}

@article{Mourad:2024mpg,
    author = "Mourad, J. and Raucci, S. and Sagnotti, A.",
    title = "{Brane profiles of non-supersymmetric strings}",
    eprint = "2406.16327",
    archivePrefix = "arXiv",
    primaryClass = "hep-th",
    doi = "10.1007/JHEP09(2024)019",
    journal = "JHEP",
    volume = "09",
    pages = "019",
    year = "2024"
}

@article{Morales:1998ux,
    author = "Morales, Jose F. and Scrucca, Claudio A. and Serone, Marco",
    title = "{Anomalous couplings for D-branes and O-planes}",
    eprint = "hep-th/9812071",
    archivePrefix = "arXiv",
    reportNumber = "SPIN-1998-13, LMU-TPW-98-17",
    doi = "10.1016/S0550-3213(99)00217-5",
    journal = "Nucl. Phys. B",
    volume = "552",
    pages = "291--315",
    year = "1999"
}

@article{McNamara:2019rup,
    author = "McNamara, Jacob and Vafa, Cumrun",
    title = "{Cobordism Classes and the Swampland}",
    eprint = "1909.10355",
    archivePrefix = "arXiv",
    primaryClass = "hep-th",
    month = "9",
    year = "2019"
}

@article{Stout:2021ubb,
    author = "Stout, John",
    title = "{Infinite Distance Limits and Information Theory}",
    eprint = "2106.11313",
    archivePrefix = "arXiv",
    primaryClass = "hep-th",
    month = "6",
    year = "2021"
}

@article{Stout:2022phm,
    author = "Stout, John",
    title = "{Infinite Distances and Factorization}",
    eprint = "2208.08444",
    archivePrefix = "arXiv",
    primaryClass = "hep-th",
    month = "8",
    year = "2022"
}

@article{Velazquez:2022eco,
    author = "Vel\'azquez, David Mart\'\i{}n and De Biasio, Davide and Lust, Dieter",
    title = "{Cobordism, singularities and the Ricci flow conjecture}",
    eprint = "2209.10297",
    archivePrefix = "arXiv",
    primaryClass = "hep-th",
    doi = "10.1007/JHEP01(2023)126",
    journal = "JHEP",
    volume = "01",
    pages = "126",
    year = "2023"
}

@article{DeBiasio:2022zuh,
    author = "De Biasio, Davide",
    title = "{On-Shell Flow}",
    eprint = "2211.04231",
    archivePrefix = "arXiv",
    primaryClass = "hep-th",
    month = "11",
    year = "2022"
}

@article{Basile:2023rvm,
    author = "Basile, Ivano and Montella, Carmine",
    title = "{Domain walls and distances in discrete landscapes}",
    eprint = "2309.04519",
    archivePrefix = "arXiv",
    primaryClass = "hep-th",
    doi = "10.1007/JHEP02(2024)227",
    journal = "JHEP",
    volume = "02",
    pages = "227",
    year = "2024"
}

@article{Li:2023gtt,
    author = "Li, Yixuan and Palti, Eran and Petri, Nicol\`o",
    title = "{Towards AdS distances in string theory}",
    eprint = "2306.02026",
    archivePrefix = "arXiv",
    primaryClass = "hep-th",
    doi = "10.1007/JHEP08(2023)210",
    journal = "JHEP",
    volume = "08",
    pages = "210",
    year = "2023"
}

@article{Palti:2024voy,
    author = "Palti, Eran and Petri, Nicol\`o",
    title = "{A positive metric over DGKT vacua}",
    eprint = "2405.01084",
    archivePrefix = "arXiv",
    primaryClass = "hep-th",
    doi = "10.1007/JHEP06(2024)019",
    journal = "JHEP",
    volume = "06",
    pages = "019",
    year = "2024"
}

@article{Mohseni:2024njl,
    author = "Mohseni, Amineh and Montero, Miguel and Vafa, Cumrun and Valenzuela, Irene",
    title = "{On measuring distances in the quantum gravity landscape}",
    eprint = "2407.02705",
    archivePrefix = "arXiv",
    primaryClass = "hep-th",
    reportNumber = "IFT-24-097, CERN-TH-2024-101",
    doi = "10.1007/JHEP12(2024)168",
    journal = "JHEP",
    volume = "12",
    pages = "168",
    year = "2024"
}

@article{Basile:2022zee,
    author = "Basile, Ivano",
    title = "{Emergent Strings at an Infinite Distance with Broken Supersymmetry}",
    eprint = "2201.08851",
    archivePrefix = "arXiv",
    primaryClass = "hep-th",
    doi = "10.3390/astronomy2030015",
    journal = "Astronomy",
    volume = "2",
    number = "3",
    pages = "206--225",
    year = "2023"
}

@article{Dvali:2017eba,
    author = "Dvali, Gia and Gomez, Cesar and Zell, Sebastian",
    title = "{Quantum Break-Time of de Sitter}",
    eprint = "1701.08776",
    archivePrefix = "arXiv",
    primaryClass = "hep-th",
    reportNumber = "LMU-ASC-08-17, MPP-2017-10",
    doi = "10.1088/1475-7516/2017/06/028",
    journal = "JCAP",
    volume = "06",
    pages = "028",
    year = "2017"
}

@article{Dvali:2018jhn,
    author = "Dvali, Gia and Gomez, Cesar and Zell, Sebastian",
    title = "{Quantum Breaking Bound on de Sitter and Swampland}",
    eprint = "1810.11002",
    archivePrefix = "arXiv",
    primaryClass = "hep-th",
    reportNumber = "LMU-ASC 71/18",
    doi = "10.1002/prop.201800094",
    journal = "Fortsch. Phys.",
    volume = "67",
    number = "1-2",
    pages = "1800094",
    year = "2019"
}

@article{Buratti:2021fiv,
    author = "Buratti, Ginevra and Calder\'on-Infante, Jos\'e and Delgado, Matilda and Uranga, Angel M.",
    title = "{Dynamical Cobordism and Swampland Distance Conjectures}",
    eprint = "2107.09098",
    archivePrefix = "arXiv",
    primaryClass = "hep-th",
    doi = "10.1007/JHEP10(2021)037",
    journal = "JHEP",
    volume = "10",
    pages = "037",
    year = "2021"
}

@article{Blumenhagen:2022mqw,
    author = "Blumenhagen, Ralph and Cribiori, Niccol\`o and Kneissl, Christian and Makridou, Andriana",
    title = "{Dynamical cobordism of a domain wall and its companion defect 7-brane}",
    eprint = "2205.09782",
    archivePrefix = "arXiv",
    primaryClass = "hep-th",
    reportNumber = "MPP-2022-57",
    doi = "10.1007/JHEP08(2022)204",
    journal = "JHEP",
    volume = "08",
    pages = "204",
    year = "2022"
}

@article{Angius:2022aeq,
    author = "Angius, Roberta and Calder\'on-Infante, Jos\'e and Delgado, Matilda and Huertas, Jes\'us and Uranga, Angel M.",
    title = "{At the end of the world: Local Dynamical Cobordism}",
    eprint = "2203.11240",
    archivePrefix = "arXiv",
    primaryClass = "hep-th",
    reportNumber = "IFT-UAM/CSIC-22-31",
    doi = "10.1007/JHEP06(2022)142",
    journal = "JHEP",
    volume = "06",
    pages = "142",
    year = "2022"
}

@article{Blumenhagen:2023abk,
    author = "Blumenhagen, Ralph and Kneissl, Christian and Wang, Chuying",
    title = "{Dynamical Cobordism Conjecture: solutions for end-of-the-world branes}",
    eprint = "2303.03423",
    archivePrefix = "arXiv",
    primaryClass = "hep-th",
    reportNumber = "MPP-2023-33",
    doi = "10.1007/JHEP05(2023)123",
    journal = "JHEP",
    volume = "05",
    pages = "123",
    year = "2023"
}

@article{Angius:2023xtu,
    author = "Angius, Roberta and Huertas, Jesus and Uranga, Angel M.",
    title = "{Small black hole explosions}",
    eprint = "2303.15903",
    archivePrefix = "arXiv",
    primaryClass = "hep-th",
    reportNumber = "IFT-UAM/CSIC-23-31",
    doi = "10.1007/JHEP06(2023)070",
    journal = "JHEP",
    volume = "06",
    pages = "070",
    year = "2023"
}

@article{Huertas:2023syg,
    author = "Huertas, Jes\'us and Uranga, Angel M.",
    title = "{Aspects of Dynamical Cobordism in AdS/CFT}",
    eprint = "2306.07335",
    archivePrefix = "arXiv",
    primaryClass = "hep-th",
    month = "6",
    year = "2023"
}

@article{Calderon-Infante:2023ler,
    author = "Calder\'on-Infante, Jos\'e and Castellano, Alberto and Herr\'aez, Alvaro and Ib\'a\~nez, Luis E.",
    title = "{Entropy Bounds and the Species Scale Distance Conjecture}",
    eprint = "2306.16450",
    archivePrefix = "arXiv",
    primaryClass = "hep-th",
    month = "6",
    year = "2023"
}

@article{Angius:2023uqk,
    author = "Angius, Roberta and Makridou, Andriana and Uranga, Angel M.",
    title = "{Intersecting End of the World Branes}",
    eprint = "2312.16286",
    archivePrefix = "arXiv",
    primaryClass = "hep-th",
    month = "12",
    year = "2023"
}

@article{Angius:2024zjv,
    author = "Angius, Roberta",
    title = "{End of The World brane networks for infinite distance limits in CY moduli space}",
    eprint = "2404.14486",
    archivePrefix = "arXiv",
    primaryClass = "hep-th",
    reportNumber = "IFT-UAM/CSIC-24-62",
    month = "4",
    year = "2024"
}

@article{Banerjee:2020wov,
    author = "Banerjee, Souvik and Danielsson, Ulf and Giri, Suvendu",
    title = "{Bubble needs strings}",
    eprint = "2009.01597",
    archivePrefix = "arXiv",
    primaryClass = "hep-th",
    reportNumber = "UUITP-33/20",
    doi = "10.1007/JHEP03(2021)250",
    journal = "JHEP",
    volume = "21",
    pages = "250",
    year = "2020"
}

@article{Cicoli:2023opf,
    author = "Cicoli, Michele and Conlon, Joseph P. and Maharana, Anshuman and Parameswaran, Susha and Quevedo, Fernando and Zavala, Ivonne",
    title = "{String cosmology: From the early universe to today}",
    eprint = "2303.04819",
    archivePrefix = "arXiv",
    primaryClass = "hep-th",
    doi = "10.1016/j.physrep.2024.01.002",
    journal = "Phys. Rept.",
    volume = "1059",
    pages = "1--155",
    year = "2024"
}

@article{VanRiet:2023pnx,
    author = "Van Riet, Thomas and Zoccarato, Gianluca",
    title = "{Beginners lectures on flux compactifications and related Swampland topics}",
    eprint = "2305.01722",
    archivePrefix = "arXiv",
    primaryClass = "hep-th",
    doi = "10.1016/j.physrep.2023.11.003",
    journal = "Phys. Rept.",
    volume = "1049",
    pages = "1--51",
    year = "2024"
}

@article{Bena:2023sks,
    author = "Bena, Iosif and Gra\~na, Mariana and Van Riet, Thomas",
    title = "{Trustworthy de Sitter compactifications of string theory: a comprehensive review}",
    eprint = "2303.17680",
    archivePrefix = "arXiv",
    primaryClass = "hep-th",
    month = "3",
    year = "2023"
}

@article{Brahma:2020tak,
    author = "Brahma, Suddhasattwa and Dasgupta, Keshav and Tatar, Radu",
    title = "{de Sitter Space as a Glauber-Sudarshan State}",
    eprint = "2007.11611",
    archivePrefix = "arXiv",
    primaryClass = "hep-th",
    doi = "10.1007/JHEP02(2021)104",
    journal = "JHEP",
    volume = "02",
    pages = "104",
    year = "2021"
}

@article{Bernardo:2021rul,
    author = "Bernardo, Heliudson and Brahma, Suddhasattwa and Dasgupta, Keshav and Faruk, Mir-Mehedi and Tatar, Radu",
    title = "{de Sitter Space as a Glauber-Sudarshan State: II}",
    eprint = "2108.08365",
    archivePrefix = "arXiv",
    primaryClass = "hep-th",
    doi = "10.1002/prop.202100131",
    journal = "Fortsch. Phys.",
    volume = "69",
    number = "11-12",
    pages = "2100131",
    year = "2021"
}

@article{Chakravarty:2024pec,
    author = "Chakravarty, Joydeep and Dasgupta, Keshav",
    title = "{What if string theory has a de Sitter excited state?}",
    eprint = "2404.11680",
    archivePrefix = "arXiv",
    primaryClass = "hep-th",
    doi = "10.1007/JHEP10(2024)065",
    journal = "JHEP",
    volume = "10",
    pages = "065",
    year = "2024"
}

@article{Demirtas:2021nlu,
    author = "Demirtas, Mehmet and Kim, Manki and McAllister, Liam and Moritz, Jakob and Rios-Tascon, Andres",
    title = "{Small cosmological constants in string theory}",
    eprint = "2107.09064",
    archivePrefix = "arXiv",
    primaryClass = "hep-th",
    doi = "10.1007/JHEP12(2021)136",
    journal = "JHEP",
    volume = "12",
    pages = "136",
    year = "2021"
}

@article{Demirtas:2021ote,
    author = "Demirtas, Mehmet and Kim, Manki and McAllister, Liam and Moritz, Jakob and Rios-Tascon, Andres",
    title = "{Exponentially Small Cosmological Constant in String Theory}",
    eprint = "2107.09065",
    archivePrefix = "arXiv",
    primaryClass = "hep-th",
    doi = "10.1103/PhysRevLett.128.011602",
    journal = "Phys. Rev. Lett.",
    volume = "128",
    number = "1",
    pages = "011602",
    year = "2022"
}

@article{McAllister:2024lnt,
    author = "McAllister, Liam and Moritz, Jakob and Nally, Richard and Schachner, Andreas",
    title = "{Candidate de Sitter Vacua}",
    eprint = "2406.13751",
    archivePrefix = "arXiv",
    primaryClass = "hep-th",
    reportNumber = "CERN-TH-2024-090",
    month = "6",
    year = "2024"
}

@article{Randall:1999ee,
    author = "Randall, Lisa and Sundrum, Raman",
    title = "{A Large mass hierarchy from a small extra dimension}",
    eprint = "hep-ph/9905221",
    archivePrefix = "arXiv",
    reportNumber = "MIT-CTP-2860, PUPT-1860, BUHEP-99-9",
    doi = "10.1103/PhysRevLett.83.3370",
    journal = "Phys. Rev. Lett.",
    volume = "83",
    pages = "3370--3373",
    year = "1999"
}

@article{Dvali:2000hr,
    author = "Dvali, G. R. and Gabadadze, Gregory and Porrati, Massimo",
    title = "{4-D gravity on a brane in 5-D Minkowski space}",
    eprint = "hep-th/0005016",
    archivePrefix = "arXiv",
    reportNumber = "NYU-TH-00-04-01",
    doi = "10.1016/S0370-2693(00)00669-9",
    journal = "Phys. Lett. B",
    volume = "485",
    pages = "208--214",
    year = "2000"
}

@article{Lust:2004ks,
    author = "Lust, Dieter",
    editor = "Bertolini, M. and Demasure, Y. and Di Vecchia, P. and Kristjansen, C. and Merlatti, P. and Obers, N.",
    title = "{Intersecting brane worlds: A Path to the standard model?}",
    eprint = "hep-th/0401156",
    archivePrefix = "arXiv",
    reportNumber = "HU-EP-04-04, MPP-2004-8",
    doi = "10.1088/0264-9381/21/10/013",
    journal = "Class. Quant. Grav.",
    volume = "21",
    pages = "S1399--1424",
    year = "2004"
}

@article{Aldazabal:2000cn,
    author = "Aldazabal, G. and Franco, S. and Ibanez, Luis E. and Rabadan, R. and Uranga, A. M.",
    title = "{Intersecting brane worlds}",
    eprint = "hep-ph/0011132",
    archivePrefix = "arXiv",
    reportNumber = "CAB-IB-2919500, CERN-TH-2000-320, MIT-CTP-3042, FTUAM-00-23, IFT-UAM-CSIC-00-37",
    doi = "10.1088/1126-6708/2001/02/047",
    journal = "JHEP",
    volume = "02",
    pages = "047",
    year = "2001"
}

@article{Antoniadis:2019rkh,
    author = "Antoniadis, Ignatios and Chen, Yifan and Leontaris, George K.",
    title = "{Logarithmic loop corrections, moduli stabilisation and de Sitter vacua in string theory}",
    eprint = "1909.10525",
    archivePrefix = "arXiv",
    primaryClass = "hep-th",
    doi = "10.1007/JHEP01(2020)149",
    journal = "JHEP",
    volume = "01",
    pages = "149",
    year = "2020"
}

@article{Blumenhagen:2022zzw,
    author = "Blumenhagen, Ralph and Brinkmann, Max and Makridou, Andriana",
    title = "{The dark dimension in a warped throat}",
    eprint = "2208.01057",
    archivePrefix = "arXiv",
    primaryClass = "hep-th",
    reportNumber = "MPP-2022-94",
    doi = "10.1016/j.physletb.2023.137699",
    journal = "Phys. Lett. B",
    volume = "838",
    pages = "137699",
    year = "2023"
}

@article{Basile:2024lcz,
    author = "Basile, Ivano and Lust, Dieter",
    title = "{Dark dimension with (little) strings attached}",
    eprint = "2409.12231",
    archivePrefix = "arXiv",
    primaryClass = "hep-th",
    month = "9",
    year = "2024"
}

@article{Anchordoqui:2025nmb,
    author = "Anchordoqui, Luis and Antoniadis, Ignatios and Lust, Dieter",
    title = "{Two Micron-Size Dark Dimensions}",
    eprint = "2501.11690",
    archivePrefix = "arXiv",
    primaryClass = "hep-th",
    reportNumber = "MPP-2025-5, LMU-ASC 02/25",
    month = "1",
    year = "2025"
}

@article{Anchordoqui:2024xvl,
    author = "Anchordoqui, Luis A. and Antoniadis, Ignatios and Lust, Dieter and Castillo, Karem Pe\~nal\'o",
    title = "{Cosmological constraints on dark neutrino towers}",
    eprint = "2411.07029",
    archivePrefix = "arXiv",
    primaryClass = "hep-ph",
    reportNumber = "MPP-2024-219; LMU-ASC 19/24",
    doi = "10.1103/PhysRevD.111.015024",
    journal = "Phys. Rev. D",
    volume = "111",
    number = "1",
    pages = "015024",
    year = "2025"
}

@article{Antoniadis:2024sfa,
    author = "Antoniadis, Ignatios and Anchordoqui, Luis A. and Lust, Dieter",
    title = "{Landscape, Swampland and extra dimensions}",
    eprint = "2405.04427",
    archivePrefix = "arXiv",
    primaryClass = "hep-th",
    doi = "10.22323/1.463.0215",
    journal = "PoS",
    volume = "CORFU2023",
    pages = "215",
    year = "2024"
}

@article{Anchordoqui:2023oqm,
    author = "Anchordoqui, Luis A. and Antoniadis, Ignatios and Cribiori, Niccol\`o and Lust, Dieter and Scalisi, Marco",
    title = "{The Scale of Supersymmetry Breaking and the Dark Dimension}",
    eprint = "2301.07719",
    archivePrefix = "arXiv",
    primaryClass = "hep-th",
    reportNumber = "MPP-2023-5, LMU-ASC 01/23",
    doi = "10.1007/JHEP05(2023)060",
    journal = "JHEP",
    volume = "05",
    pages = "060",
    year = "2023"
}

@article{Anchordoqui:2022txe,
    author = "Anchordoqui, Luis A. and Antoniadis, Ignatios and Lust, Dieter",
    title = "{Dark dimension, the swampland, and the dark matter fraction composed of primordial black holes}",
    eprint = "2206.07071",
    archivePrefix = "arXiv",
    primaryClass = "hep-th",
    reportNumber = "MPP-2022-60, LMU-ASC 24/22",
    doi = "10.1103/PhysRevD.106.086001",
    journal = "Phys. Rev. D",
    volume = "106",
    number = "8",
    pages = "086001",
    year = "2022"
}

@article{Anchordoqui:2022tgp,
    author = "Anchordoqui, Luis A. and Antoniadis, Ignatios and Lust, Dieter",
    title = "{The dark universe: Primordial black hole \ensuremath{\leftrightharpoons} dark graviton gas connection}",
    eprint = "2210.02475",
    archivePrefix = "arXiv",
    primaryClass = "hep-th",
    reportNumber = "MPP-2022-125, LMU-ASC 29/22",
    doi = "10.1016/j.physletb.2023.137844",
    journal = "Phys. Lett. B",
    volume = "840",
    pages = "137844",
    year = "2023"
}

@article{Vafa:2025nst,
    author = "Vafa, Cumrun",
    title = "{On the origin and fate of our universe}",
    eprint = "2501.00966",
    archivePrefix = "arXiv",
    primaryClass = "hep-th",
    doi = "10.1007/s10714-025-03353-w",
    journal = "Gen. Rel. Grav.",
    volume = "57",
    number = "1",
    pages = "19",
    year = "2025"
}

@article{Vafa:2024fpx,
    author = "Vafa, Cumrun",
    title = "{Swamplandish Unification of the Dark Sector}",
    eprint = "2402.00981",
    archivePrefix = "arXiv",
    primaryClass = "hep-ph",
    month = "2",
    year = "2024"
}

@article{Obied:2023clp,
    author = "Obied, Georges and Dvorkin, Cora and Gonzalo, Eduardo and Vafa, Cumrun",
    title = "{Dark dimension and decaying dark matter gravitons}",
    eprint = "2311.05318",
    archivePrefix = "arXiv",
    primaryClass = "astro-ph.CO",
    doi = "10.1103/PhysRevD.109.063540",
    journal = "Phys. Rev. D",
    volume = "109",
    number = "6",
    pages = "063540",
    year = "2024"
}

@article{Gonzalo:2022jac,
    author = "Gonzalo, Eduardo and Montero, Miguel and Obied, Georges and Vafa, Cumrun",
    title = "{Dark dimension gravitons as dark matter}",
    eprint = "2209.09249",
    archivePrefix = "arXiv",
    primaryClass = "hep-ph",
    doi = "10.1007/JHEP11(2023)109",
    journal = "JHEP",
    volume = "11",
    pages = "109",
    year = "2023"
}

@article{Coudarchet:2023mfs,
    author = "Coudarchet, Thibaut",
    title = "{Hiding the extra dimensions: A review on scale separation in string theory}",
    eprint = "2311.12105",
    archivePrefix = "arXiv",
    primaryClass = "hep-th",
    doi = "10.1016/j.physrep.2024.02.003",
    journal = "Phys. Rept.",
    volume = "1064",
    pages = "1--28",
    year = "2024"
}

@article{Delgado:2024skw,
    author = "Delgado, Matilda and van de Heisteeg, Damian and Raman, Sanjay and Torres, Ethan and Vafa, Cumrun and Xu, Kai",
    title = "{Finiteness and the Emergence of Dualities}",
    eprint = "2412.03640",
    archivePrefix = "arXiv",
    primaryClass = "hep-th",
    reportNumber = "MPP-2024-224, CERN-TH-2024-204",
    month = "12",
    year = "2024"
}

@article{Kim:2024hxe,
    author = "Kim, Hee-Cheol and Vafa, Cumrun and Xu, Kai",
    title = "{Finite Landscape of 6d N=(1,0) Supergravity}",
    eprint = "2411.19155",
    archivePrefix = "arXiv",
    primaryClass = "hep-th",
    month = "11",
    year = "2024"
}

@article{Montero:2020icj,
    author = "Montero, Miguel and Vafa, Cumrun",
    title = "{Cobordism Conjecture, Anomalies, and the String Lamppost Principle}",
    eprint = "2008.11729",
    archivePrefix = "arXiv",
    primaryClass = "hep-th",
    doi = "10.1007/JHEP01(2021)063",
    journal = "JHEP",
    volume = "01",
    pages = "063",
    year = "2021"
}

@article{Bedroya:2021fbu,
    author = "Bedroya, Alek and Hamada, Yuta and Montero, Miguel and Vafa, Cumrun",
    title = "{Compactness of brane moduli and the String Lamppost Principle in d \ensuremath{>} 6}",
    eprint = "2110.10157",
    archivePrefix = "arXiv",
    primaryClass = "hep-th",
    doi = "10.1007/JHEP02(2022)082",
    journal = "JHEP",
    volume = "02",
    pages = "082",
    year = "2022"
}

@article{Hamada:2021yxy,
    author = "Hamada, Yuta and Montero, Miguel and Vafa, Cumrun and Valenzuela, Irene",
    title = "{Finiteness and the swampland}",
    eprint = "2111.00015",
    archivePrefix = "arXiv",
    primaryClass = "hep-th",
    doi = "10.1088/1751-8121/ac6404",
    journal = "J. Phys. A",
    volume = "55",
    number = "22",
    pages = "224005",
    year = "2022"
}

@article{VanRaamsdonk:2024sdp,
    author = "Van Raamsdonk, Mark and Waddell, Chris",
    title = "{Holographic motivations and observational evidence for decreasing dark energy}",
    eprint = "2406.02688",
    archivePrefix = "arXiv",
    primaryClass = "hep-th",
    month = "6",
    year = "2024"
}

@inproceedings{Johnson:2000ch,
    author = "Johnson, Clifford V.",
    title = "{D-brane primer}",
    booktitle = "{Theoretical Advanced Study Institute in Elementary Particle Physics (TASI 99): Strings, Branes, and Gravity}",
    eprint = "hep-th/0007170",
    archivePrefix = "arXiv",
    reportNumber = "DTP-00-61",
    doi = "10.1142/9789812799630_0002",
    pages = "129--350",
    month = "7",
    year = "2000"
}

\end{document}